# Oblique helicoidal state of the twist-bend nematic doped by chiral azo-compound

Vitalii Chornous[a], Alina Grozav[a], Myhaylo Vovk[b], Daria Bratova[c,d], Natalia Kasian[e], Longin Lisetski[e*], Igor Gvozdovskyy[d]*

[a]Department of Medical and Pharmaceutical Chemistry, Bukovinian State Medical University, Chernivtsi, Ukraine; [b]Department of Mechanisms of Organic Reactions, Institute of Organic Chemistry of the National Academy of Sciences of Ukraine, Kyiv, Ukraine; [c]Department of Information and Measuring Technologies, National Technical University of Ukraine "Igor Sikorsky Kyiv Polytechnic Institute"; [d]Department of Optical Quantum Electronics, Institute of Physics of the National Academy of Sciences of Ukraine, Kyiv, Ukraine; [e]Department of Nanostructured Materials, Institute for Scintillation Materials of STC "Institute for Single Crystals" of the National Academy of Sciences of Ukraine, Kharkiv, Ukraine.

Institute of Physics of the National Academy of Sciences of Ukraine, 46 Nauky ave., Kyiv, 03028, Ukraine, telephone number: +380 44 5250862, *E-mail: igvozd@gmail.com

# Oblique helicoidal state of the twist-bend nematic doped by chiral azo-compound

A novel light-sensitive chiral dopant ChD-3816 (an azo compound containing 4-hexanoyloxyphenyl and 2-isopropyl-5-methylcyclohexylbenzoate moieties) was used for inducing helical twisting in $N_{tb}$-forming mixtures of CB7CB/CB6OCB with 5CB added to decrease the phase transition temperatures. The effects of ChD-3816 upon phase transition temperatures, as well as effects of its concentration on the measured values of helical twisting were determined. Most of the measured parameters could be varied due to light-induced *trans-cis* isomerization of ChD-3816. Under electric field, selective reflection spectra in the visible range were obtained for the emerging $Ch_{OH}$ structures, with the wavelengths controllable both by electric field and appropriate UV irradiation. Possible applications for dynamic formation of contrast images are discussed.

## 1. Introduction

Though liquid crystals as a new state of matter were discovered in the course of studies of cholesterol esters about 160 years ago, [1-4] the search for new cholesteric liquid crystals (CLC) still remains relevant due to their wide practical applications. Nowadays, CLCs are no more directly related to cholesterol esters or other steroid compounds, but are typically mixtures consisting of a host nematic liquid crystal (LC) which, upon dissolution of a chiral dopant (ChD) therein, is self-reassembled forming a helicoidal structure (Figure 1(a)). In this structure, generally known as a chiral nematic ($N^*$) phase, the molecules align perpendicularly to the cholesteric helical axis. [5] The helicoidal structure is characterized by the pitch $P_0$, which depends on the helical twisting power (HTP, $\beta$) of ChDs and their concentration $C$ in nematic LCs. It can be expressed as $\beta = (P_0 \times C)^{-1}$. [6,7] Thus, for the cholesteric phase induced by a ChD with concentration $C$, the wave number is written as $q_0 = \pm 4\pi \times \beta \times C$, where $\beta$ is the efficiency of the dopant to induce a cholesteric phase with the helical pitch $P_0 = 2\pi/|q_0|$. The helical structure can be right-handed ($q_0 > 0$) or left-handed

($q_0 < 0$) depending on both the nature of ChDs molecules and their interaction with nematic LC molecules. [5]

The search for novel cholesteric and other helically twisted liquid crystal systems led to many interesting developments with promising practical applications. Thus, in recent years great attention was paid to nematic hosts with very low values of the bend elastic constant $K_{33}$. This interest was partially based on theoretical predictions of specific cholesteric-like states based on LCs with the bend elastic constant $K_{33}$ being much lower than the twist elastic constant $K_{22}$. [8,9] Later, it was assumed [10] that banana-shaped mesogenic molecules, inducing a local bend of the nematic director, could give rise to a phase with negative bend elastic constant, predicting a lower symmetry nematic phase characterised by symmetry breaking transition from uniform texture to spontaneous periodic distortion, either oscillating splay-bend or conical twist-bend helix. Another tentative model and mechanism describing possible cholesteric helix distortions of the "oblique helicoidal" nature caused by ChDs of a specific nature (*e.g.* small anisotropy and significant molecular biaxiality) was proposed, considering the relationship between elastic deformation fields and the director of quasi-nematic layer. [11]

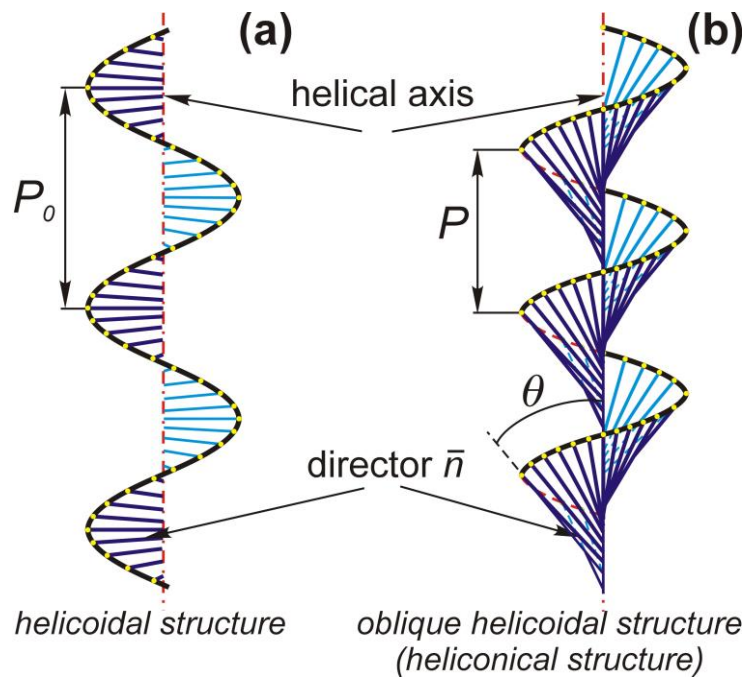

Figure 1. Schematic illustration of the structure in which local orientation of LC molecules (director $\bar{\mathbf{n}}$) rotates around the helical axis: (a) perpendicularly in the absence of electric fields for helicoidal cholesterics (N*) with the helix pitch length $P_0$,

and (b) with an oblique angle $\theta$ for heliconical cholesterics (Ch$_{OH}$) under electric field $E$. The length of the heliconical helix pitch $P$ and oblique angle $\theta$ increase with decreasing applied electric field $E$.

In this work, we used recently described twist-bend nematic (N$_{tb}$) LCs, [12,13] with a low value of bend constant K$_{33}$ to obtain chiral twist-bend nematics (N$^*_{tb}$) by adding a small quantity of ChD. [14,15]

As it could be expected, we have experimentally found that our LC mixtures possessing the N$^*_{tb}$ phase can form, at temperatures above the N$^*_{tb}$ − N$^*$ phase transition, a specific cholesteric state under AC electric field known as the heliconical structure of cholesteric (or Ch$_{OH}$ for short). [14,15] In this state, the molecules are twisted around the cholesteric helical axis with a certain oblique angle of $\theta$ (Figure 1 (b)), as distinct from the conventional CLCs, where $\theta = 90°$ [5,9]. Under applied electric field larger then the threshold field ($E_{NC}$) the oblique helicoidal structure (Figure 1 (b)) undergoes the transformation to the homeotropically aligned nematic. This threshold field can be described as follows: [8,14]

$$E_{NC} = \frac{2\pi \cdot K_{22}}{P_0 \cdot \sqrt{\varepsilon_0 \cdot \Delta\varepsilon \cdot K_{33}}}, \quad (1)$$

where $\varepsilon_0$ is the constant of vacuum permittivity and $\Delta\varepsilon$ is the dielectric anisotropy of the LC mixture, $K_{22}$ and $K_{33}$ are twist and bend elastic constants, respectively; $P_0$ is the length of cholesteric pitch of N$^*$ phase of the N$^*_{tb}$-forming mixture in the absence of applied electric field.

The Ch$_{OH}$ state can show selective reflection of light (so-called Bragg diffraction) in the visible spectral range under certain conditions, namely temperature $T_B$ and an electric field $E_B$. The maximum of wavelength $\lambda_{max}$ of the reflected light can be switched by changing of temperature, applied voltage and frequency of the electric field. [14-20] Besides, when the electric field $E_B$ is decreased, there is also a threshold field ($E_{N^*C}$), when the oblique helicoidal structure (Figure 1 (b)) transforms to conventional helicoidal structure (*i.e.* $\theta = 0°$), which is reflected by the appearance of the focal conic texture leading to intense light scattering. This threshold electric field can be expressed as:

$$E_{N*C} \approx E_{NC} \cdot \frac{K_{33}}{K_{22}+K_{33}}[2+\sqrt{2\cdot(1-\frac{K_{33}}{K_{22}})}] \qquad (2)$$

Thus, the main condition for the existence of Ch$_{OH}$ structure is a non-zero applied electric field $E$ at a certain temperature $T$ (*i.e.* when $K_{33}/K_{22} <$ ½ [9]), when the following inequality is satisfied:

$$E_{N*C} \leq E \leq E_{NC} \qquad (3)$$

However, it should be noted that in reality the Ch$_{OH}$ structure will display the selective Bragg reflection of light in the visible range in a somewhat narrower interval of electrical field $E$, than expressed in the Inequality (3), and can be written as follows:

$$E_B^{\lambda_2} \leq E \leq E_B^{\lambda_1}, \qquad (4)$$

where $E_B^{\lambda_1}$ and $E_B^{\lambda_2}$ are values of electric fields $E$ under which the Bragg reflection of light is observed in the visible range ($\lambda_1 < \lambda_2$).

The detailed studies of the oblique helicoidal cholesterics and their applications are described in Refs [14-23].

The idea to use photosensitive chiral dopants to obtain N*$_{tb}$-forming mixtures that, above the N*$_{tb}$ − N* transition display the oblique helicoidal cholesteric state under applied field was recently realized. [21-23] The oblique helicoidal cholesterics consisting of both non-chiral azoxybenzene derivative [21,22] and a chiral photosensitive compound [23] were characterized by an additional useful property, namely, photo-controllable switching of selective reflected wavelength.

The N$_{tb}$-forming systems are usually characterized by a sequence of twist-bend nematic − nematic − isotropic (N$_{tb}$ − N − Iso) phase transitions. [16,17,19] As it was theoretically predicted in [24] the adding of chiral and achiral dopants to such systems should affect the phase diagrams and helical twisting. Experimentally it was shown that the decreasing of the temperatures of phase transitions is also obtained when adding both the nematic 5CB and different ChDs. [25]

In this manuscript we describe chiral N$_{tb}$-based phases doped with light-sensitive chiral dopant in various concentrations that display the oblique helicoidal cholesteric state under applied electric field. The influence of the concentration of light-sensitive chiral dopant and irradiation time on temperatures of phase transitions of $N^*_{tb}$ and the magnitude of the electric field needed to switch the wavelength of the Bragg reflection in oblique helicoidal cholesteric state will be described.

## 2. Experiment

To obtain a $N_{tb}$-forming system with low temperature phase transitions, we used the three-component nematic mixture containing two achiral liquid crystals dimers, namely 1'-,7''-bis-4-(4-cyanobiphen-4'-yl)heptane (CB7CB) and 1-(4-cyanobiphenyl-4'-yloxy)-6-(4-cyanobiphenyl-4'-yl)hexane (CB6OCB) and monomer nematic 4-pentyl-4'-cyanobiphenyl (K15 or 5CB) in the ratio of (39:19:42). Both twist-bend nematics were obtained from Synthon Chemicals GmbH & Co (Wolfen, Germany), while the nematic 5CB was synthesized and purified before using at the STC "Institute of Single Crystals" (Kharkiv, Ukraine). It is known that CB7CB possesses a uniaxial nematic N phase in a temperature range of 103 - 116 ºC, [12,26,27] with the elastic constants $K_{11} \approx 8$ pN, $K_{22} \approx 2.5$ pN and $K_{33} \approx 0.39$ pN measured at 102.3 ºC, [12] while CB6OCB possesses a uniaxial N phase in a temperature range of 110 - 157 ºC. [27-29] The temperature dependence of the elastic constants twist-bend nematic CB7CB was measured in Ref. [30] Chemical structures of the nematic molecules forming the basic $N_{tb}$ mixture are shown in Figure 2 (a).

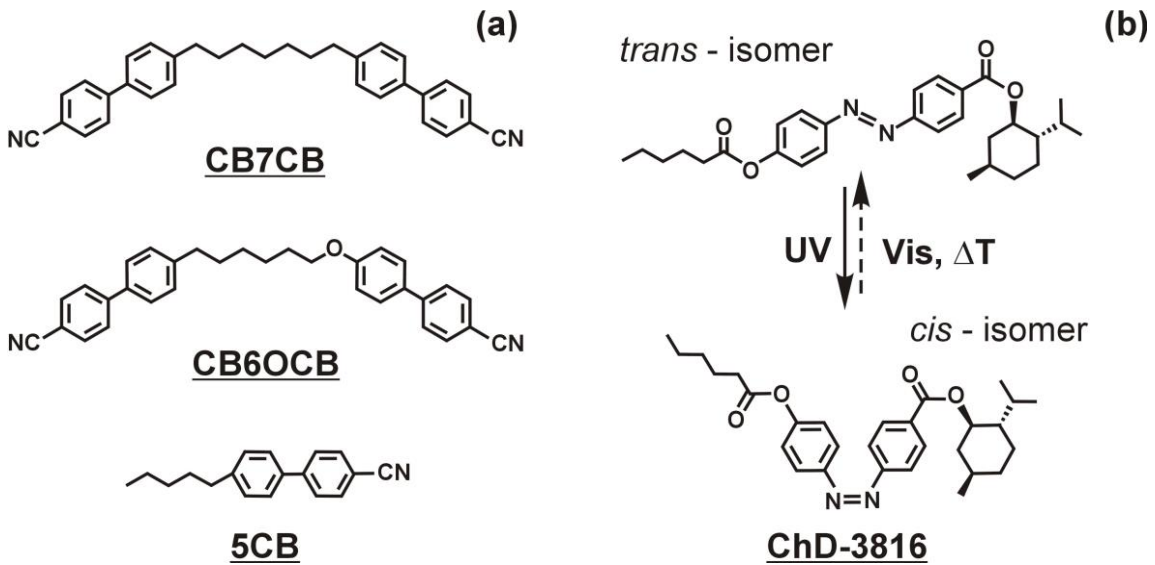

Figure 2. Schematic image of molecular structures of the nematics: CB7CB, CB6OCB and 5CB (a) and the light-sensitive chiral dopant ChD-3816 in *trans*- and *cis*-forms (b).

To induce chirality in the twist-bend nematic, we used (1*R*,2*S*,5*R*)-2-isopropyl-5-methylcyclohexyl-4-{(*E*)-[4-(hexanoyloxy)phenyl]diazenyl}benzoate (or ChD-3816 for short) synthesized in Bukovinian State Medical University (Chernivtsi, Ukraine) as a left-handed light-sensitive chiral dopant (Supporting Information). Chemical structures of the *trans*- and *cis*-isomers of the ChD-3816 molecule are shown in Figure 2 (b).

To study the influence of the chiral dopant on the properties of the emerging LC phases, the three-component nematic mixture was doped by ChD-3816 within concentration range 6 - 13 wt. %.

To obtain planar alignment, *n*-methyl-2-pyrrolidone solution of the polyimide PI2555 (HD MicroSystems, USA) [31] in proportion 10:1 was used. PI2555 films were spin-coated (6800 rpm, 10 s) on glass substrates (microscope slides, made in Germany) covered by indium tin oxide (ITO) layer. Substrates were dried at 80 ºC for 15 min, followed by annealing at 180 ºC for 30 min. The thin polyimide layers were unidirectionally rubbed $N_{rubb}$ = 15 times to provide their strong azimuthal anchoring energy. [32]

To measure the length of cholesteric pitch $P_0$ of the N$^*$ phase, the Grandjean-Cano method [5] was used. The wedge-like LC cells were assembled by using pair glass substrates covered with PI2555 film having opposite rubbing directions and dimensions $10 \times 15$ mm$^2$. The thickness of cells $d$ was set to 3 - 10 μm by Mylar spacer.

The plane-parallel LC cells were assembled with a thickness set by 20 μm diameter Mylar spacer and controlled by the interference method, measuring the transmission spectrum of the empty cell by means of the spectrometer Ocean Optics USB4000 (Ocean Insight, USA, California). LC cells were filled by capillary action in the isotropic phase (Iso) of N$^*_{tb}$ and further slowly cooled.

To observe the selective Bragg reflection of light AC voltage within the range 0 - 100 V with a frequency 1 kHz was applied to samples at constant temperature within range 27 - 44 °C.

For irradiation of LC mixtures containing the light-sensitive ChD-3816, the UV lamp with $\lambda_{max}$ = 365 nm was used. The illumination of LC cells was carried out though masks of arbitrary shape.

The temperatures of phase transitions of the mixtures containing various concentration of ChD-3816 were studied in a thermostable heater based on a temperature regulator MikRa 603 (LLD 'MikRa', Kyiv, Ukraine) equipped with a platinum resistance thermometer Pt1000 (PJSC 'TERA', Chernihiv, Ukraine). The temperature measurement accuracy was ± 0.1 °C/min.

The phase textures were observed by means of the polarising optical microscope (POM) BioLar PI (Warsaw, Poland) equipped by digital camera Nikon D80.

## 3. Results and discussions

*3.1. Phase transitions of nematic/chiral nematic mixtures with the low-temperature twist-bend phase*

In this section, we discuss the effects of the light-sensitive chiral dopant ChD-3816 containing azo-fragment on the temperatures of $N_{tb}$ – N – Iso phase transitions of the CB7CB:CB6OCB:5CB mixture.

In case of the $N_{tb}$ phase formed by flexible achiral dimers like CB7CB, the temperature of phase transition can be lowered by adding a certain concentration of nematic 5CB consisting of rod-like molecules [33]. Recently it was shown that for $N_{tb}$ mixture consisting of two flexible achiral dimers CB7CB and CB6OCB in ratio (1:1) the temperatures of phase transitions were decreased upon adding both the nematic 5CB and different ChDs (*e.g.* R-811 and cholesteryl oleyl carbonate). [25]

As shown in Ref [15], three-component nematic mixtures CB7CB:CB6OCB:5CB in certain ratios, when doped with a small concentration of left-handed ChD S-811, possess the $N^*$ phase in a broad temperature range including near-room temperatures, and under applied electric field the heliconical cholesteric phase $Ch_{OH}$ is observed. Basing on these studies, we prepared an $N_{tb}$-forming mixture CB7CB:CB6OCB:5CB in weight ratio (39:19:42) doped with various concentrations of light-sensitive ChD-3816, with the aim of obtaining the $Ch_{OH}$ state under application of an electric field in a broad temperature range.

Figure 3 shows the diagram of phase transitions of the basic $N_{tb}$ mixture CB7CB:CB6OCB:5CB in weight ratio (39:19:42) during heating and cooling processes.

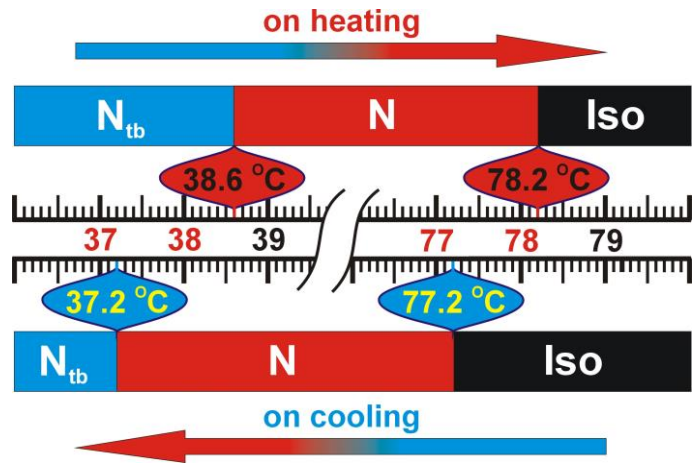

Figure 3. Schematic diagram of the sequence transitions on heating and cooling of: (a) basic $N_{tb}$ phase, containing of CB7CB, CB6OCB and 5CB in a ratio (39:19:42). The rates of the temperature change during the heating/cooling process was 0.1 °C/min.

The sequential textures of the basic $N_{tb}$ mixture during heating process in polarising optical microscope (POM) are shown in Figure 4. The $N_{tb}$ phase exists up to about 38 °C and the typical stripes texture is observed (Figure 4 (a), (b)). The $N_{tb}$ phase, placed between crossed polariser (P) and analyser (A), is characterized by birefringence as a typical nematic LCs. Figure 4 (a) shows the $N_{tb}$ phase when the rubbing direction coincides with the plane of the polarisation of P and no transmission of light is observed. The rotation of the sample by 45° leads to an increase in the sample transmittance owing to birefringence (Figure 4 (b)). The planar texture of the N phase (Figure 4 (d)) is observed in the wide temperature range about 40 °C (Figure 3)

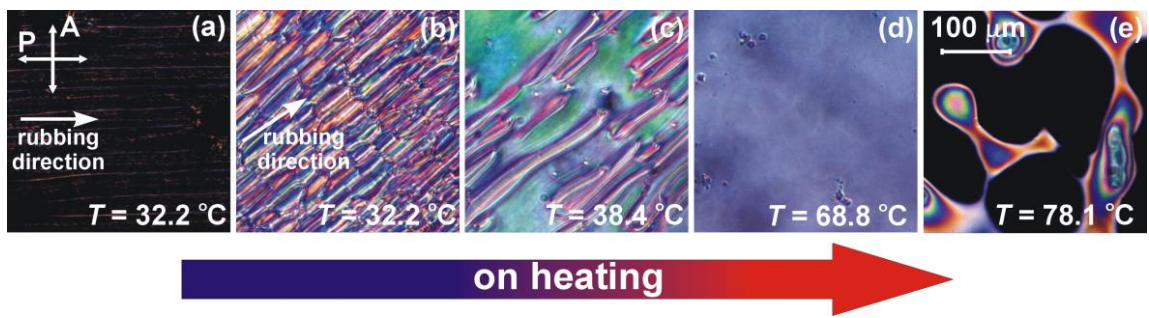

Figure 4. Photographs of the textures of the basic $N_{tb}$ mixture containing 39 wt. % CB7CB, 19 wt. % CB6OCB and 42 wt. % 5CB in 21.4 μm LC cell on heating: (a) $N_{tb}$ phase at 32.2 °C when plane polarization of polarizer P coincides with rubbing direction; (b) $N_{tb}$ phase at 32.2 °C with the 45° angle between the polarization plane of polarizer P and rubbing direction; (c) $N_{tb}$ − N phase transition at 38.4 °C; (d) N phase at

68.8 °C, and (e) N − Iso phase transition at 78.1 °C. LC cell was placed between POM crossed polarizer (P) and analyzer (A). The rubbing direction of PI2555 layers (indicated by arrow) is rotated by about 45° with respect to the polarizers.

It was experimentally found that adding certain concentration of the left-handed chiral dopant ChD-3816 within range 6 - 13 wt. % to the basic $N_{tb}$ gives rise to the formation $N^*_{tb}$, possessing the $Ch_{OH}$ state under application of an electric field.

It could be expected that, in the same way as in [25], the lowering of the phase transition temperatures of the basic $N_{tb}$ mixture would be realized by increasing the concentration of ChD-3816. The obtained results are presented in Table 1.

Table 1. Temperatures of phase transitions of the $N_{tb}$-forming mixture CB7CB:CB6OCB:5CB in weight ratio (39:19:42) containing different concentrations of ChD-3816 under heating and cooling.

| Concentration of ChD-3816, wt. % | $T_{N_{tb}/N^*_{tb}-N/N^*}$, °C | | $T_{N/N^*-Iso}$, °C | |
|---|---|---|---|---|
| | on heating | on cooling | on heating | on cooling |
| 0 | 38.6 | 37.2 | 78.2 | 77.2 |
| 6.4 | 32 | 32.2 | 62.9 | 63.1 |
| 8 | 27.5 | 27.8 | 59.1 | 59.3 |
| 10.1 | 25.2 | 25.8 | 56.9 | 57.1 |
| 12.2 | 24.3 | 25.2 | 52.7 | 53.1 |
| 13.2 | 22.8 | 23.5 | 48.2 | 48.3 |

Figure 5 shows the sequence of textures during heating process of the chiral twist-bend phase $N^*_{tb}$ based on basic $N_{tb}$ doped with 8 wt. % ChD-3816. The blocky texture of $N^*_{tb}$ phase is shown in Figure 5 (a). Due to the presence of chiral molecules, the $N^*_{tb} - N^*$ phase transition occurs at lower temperatures than with undoped $N_{tb}$ phase (Table 1). Before the $N^*_{tb} - N^*$ transition the polygonal texture appears (Figure 5(b)). For the $N^*$ phase, the Grandjean-Cano (planar) texture (with oily streaks typical for cholesteric textures) is observed (Figure 5 (c)).

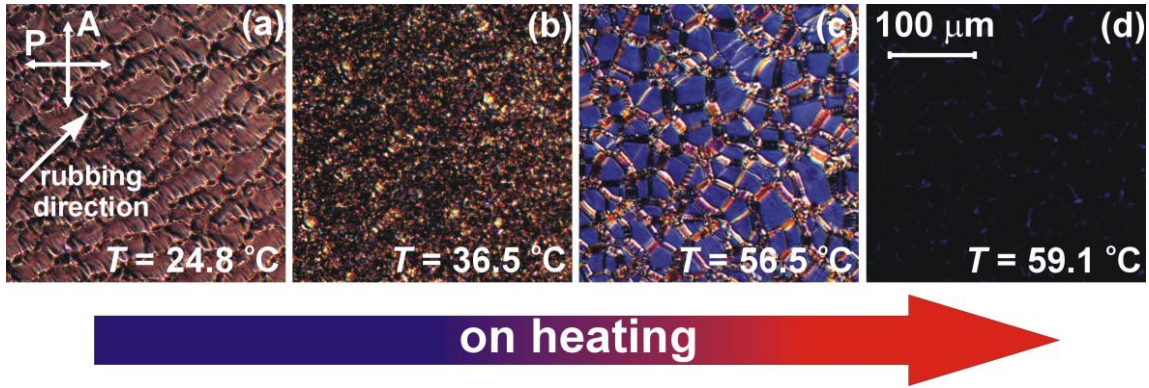

Figure 5. Photographs of the textures of the $N^*_{tb}$ phase containing the basic $N_{tb}$ mixture doped with 8 wt. % ChD-3816 during heating process: (a) blocky texture of the $N^*_{tb}$ phase at 24.8 °C; (b) polygonal texture of the $N^*_{tb}$ phase at 36.5 °C; (c) Grandjean-Cano texture and oily streaks of the $N^*$ phase; (d) $N^*$ − Iso phase transition at 59.1 °C. LC cell was placed between crossed polarizer (P) and analyzer (A) of POM. The rubbing direction (indicated by arrow) of PI2555 layers is rotated by about 45° with respect to the polarizers. Thickness of the LC cell was 20.8 μm. LC cell was placed between crossed polarizer (P) and analyzer (A) of POM. The rubbing direction (indicated by arrow) of PI2555 layers is rotated by about 45° with respect to the polarizers.

Dependence of the phase transition temperatures on the ChD-3816 concentration is shown in Figure 6(a).

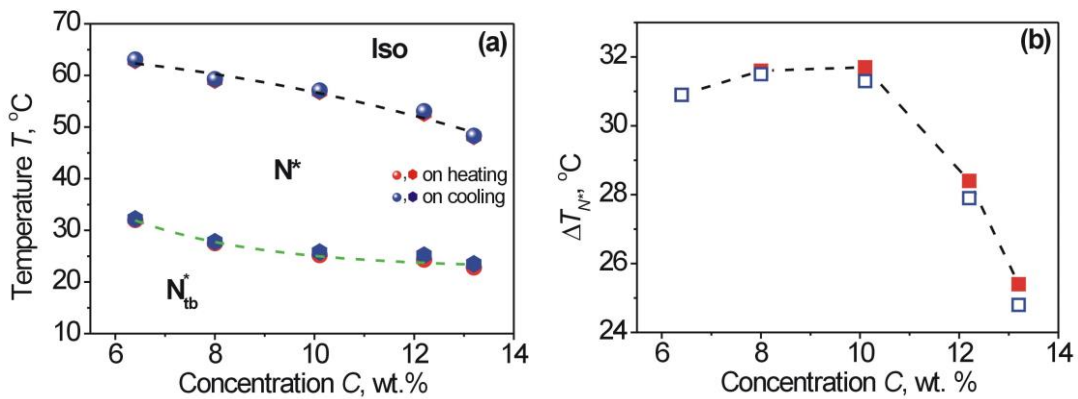

Figure 6. Dependence of the temperatures of phase transitions $N^*_{tb}$ − $N^*$ and $N^*$ − Iso (a) and the temperature range of the $N^*$ phase (b) on concentration of the ChD-3816 under heating (red symbols) and cooling (blue symbols).

We see that the phase transition temperatures are decreasing upon addition of ChD-3816. A small hysteresis on heating and cooling of $N^*_{tb}$ is observed. The $N^*$ phase exists in the wide temperature range about 25 – 32 ºC, depending on concentration of ChD-3816 both on heating and cooling. The widest possible range of temperatures of existence of the $N^*$ phase ($\Delta T_{N^*}$) is observed for concentration range 8 - 10 wt. % of ChD-3816 (Figure 6(b)), which could be important for various applications.

*3.2. Oblique helicoidal cholesterics containing the light-sensitive ChD-3816*

With all tested concentrations $C$ of ChD-3816 in the basic $N_{tb}$ mixture, no selective Bragg reflection of light could be observed in the $N^*$ phase without the applied electric field; and the helical pitch $P_0$ is a bit too high to detect wavelength $\lambda_{max}$ in the visible range. Figure 7 (a) shows the measured cholesteric helix pitch $P_0$ as function of concentration of ChD-3816 at 34 ºC (*i.e.* at this temperature, the $N^*$ phase is observed for all studied concentrations of ChD-3816 given in Table 1).

No noticeable changes of $P_0$ with temperature were noted in the entire range of the existence of $N^*$ phase. The decrease of the length of cholesteric pitch $P_0$ in the $N^*$ phase with increasing concentration of the ChD-3816 happens without selective Bragg reflection of light in visible spectrum similarly to the conventional induced cholesterics formed by small concentration of chiral dopant (*e.g.* see Figure 3S1 (a) in Supporting Information for the cholesteric mixture containing nematic E7 and the same ChD-3816 as in our study).

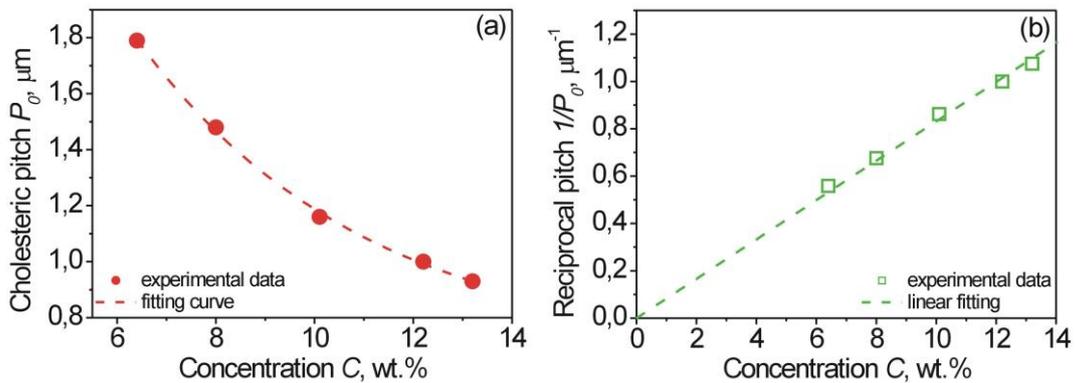

Figure 7. Dependence of cholesteric pitch $P_0$ (a) and reciprocal cholesteric pitch $1/P_0$ (b) on concentration $C$ of the ChD-3816 in the $N_{tb}$-forming mixture

CB7CB:CB6OCB:5CB in a ratio (39:19:42) in N* phase at 33 ºC in the absence of the applied electric field.

Figure 7 (b) shows the linear dependence of the reciprocal cholesteric pitch $1/P_0$ on concentration of ChD-3816. As the linear dependence $1/P_0(C)$ passes though the origin of coordinates, the value HTP ~ 0.082 (μm × wt. %)$^{-1}$ of ChD-3816 in basic $N_{tb}$ host was calculated. Thus, the HTP value of ChD-3816 in the basic $N_{tb}$ host is slightly smaller than in the nematic host E7 (Supporting Information, Figure 3S2).

It is quite obvious that within the temperature range $\Delta T_{N*}$ (Figure 6(b)) the oblique helicoidal cholesteric state displaying selective Bragg reflection of light in the visible spectral range appears at the critical applied field $E_B$, which is below the threshold electric field $E_{NC}$ and above $E_{N*C}$. To obtain the temperature range $\Delta T_{ChOH}$ of the Ch$_{OH}$ state, the measurements of the critical applied voltage $U_B$ (or electric field $E_B = U_B/d$) under switching at two wavelengths (i.e. $\lambda_1 = 430$ and $\lambda_2 = 720$ nm) for various temperatures $T$, were carried out.

Figure 8 shows the dependence of the critical electric field $E_B$ on temperature $T$ of the Ch$_{OH}$ at different concentrations of ChD-3816 in the basic $N_{tb}$ mixture. The increasing of the Ch$_{OH}$ temperature leads to the switching-on of the Bragg reflection at higher $E_B$. The values of the critical electric field $E_B^{\lambda_1=430}$ and $E_B^{\lambda_2=720}$ required to switch on the Bragg reflection increase with increasing temperature of the Ch$_{OH}$ phase. Depending on the chiral dopant concentration $C$, the value of critical electric field $E_B$ increases.

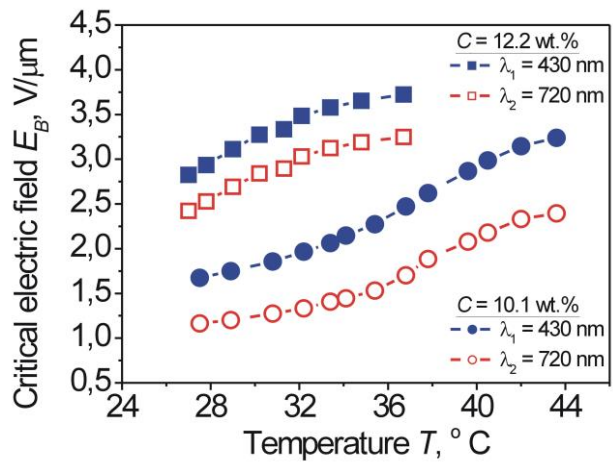

Figure 8. Dependence of the critical electric field $E_B$ on the temperature $T$ of the Ch$_{OH}$ based on N$_{tb}$ mixture (CB7CB:CB6OCB:5CB) with weight ratio (39:19:42) and ChD-3816 with concentrations: 10.1 wt. % (circles symbols) and 12.2 wt. % (squares symbols). Critical electric field value $E_B$ is at: $\lambda_1$ = 430 nm (solid blue symbols) and $\lambda_2$ = 720 nm (opened red symbols). The thickness of the LC cell was 20.5 μm (circles symbols) and 20.1 μm (squares symbols), respectively.

It should be noted that the range of the applied field $\Delta E_B$ (*i.e.* $\Delta E_B = E_B^{\lambda_1=430} - E_B^{\lambda_2=720}$) required to switch on the selective reflection in the range of the visible light spectrum also depends on both the concentration of ChD-3816 and the temperature $T$ of Ch$_{OH}$.

Figure 9 shows the dependence of the range of critical electric field $\Delta E_B$ on temperature $T$ of the Ch$_{OH}$ with various concentrations of ChD-3816. The increasing Ch$_{OH}$ temperature $T$ leads to a slight increase in the range of $\Delta E_B$, which is in a good agreement with results reported in [17]. The main cause of this is the increase in the bend elastic constants $K_{33}$ of twist-bend nematic CB7CB, as, e.g., measured experimentally in [30] Thus, we may assume that both for basic N$_{tb}$ and for N$^*_{tb}$ mixtures doped with various concentrations of ChD-3816, the bend elastic constant $K_{33}$ increases with temperature.

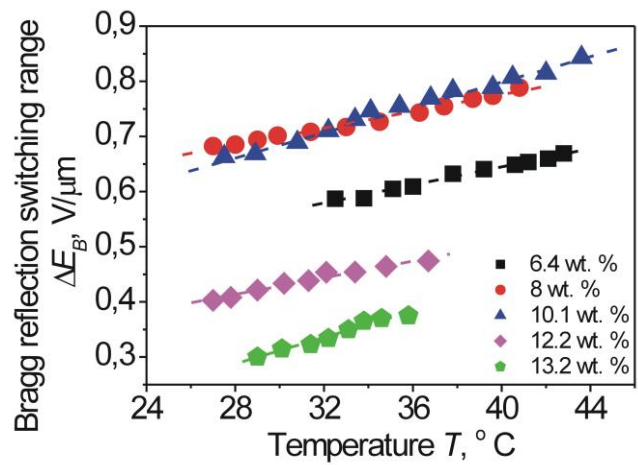

Figure 9. Dependence of the range of critical electric field $\Delta E_B$ on temperature of the Ch$_{OH}$ for ChD-3816 concentration: 6.4 wt. % (solid black squares); 8 wt. % (solid red circles); 10.1 wt. % (solid blue triangles); 12.2 wt. % (solid magenta diamonds); 12.2

wt. % (solid green pentagons).

Figure 10 shows both the range of critical electric field $\Delta E_B$ for 34 °C (a) and temperature interval $\Delta T_{ChOH}$ (b) of the $Ch_{OH}$ state where Bragg reflection of light is observed within the visible range for various concentrations of the ChD-3816. The widest range of critical electric field $\Delta E_B$ is observed for concentrations within the range about 8 - 10 wt. % (Figure 10 (a)). The widest temperature range is about 16 °C for the $Ch_{OH}$ containing 10 wt. % ChD-3816 (Figure 10 (b)). We can conclude that the $N^*_{tb}$-forming mixture with 8 - 10 wt % of the ChD3816 may become the most preferred choice option for many possible applications.

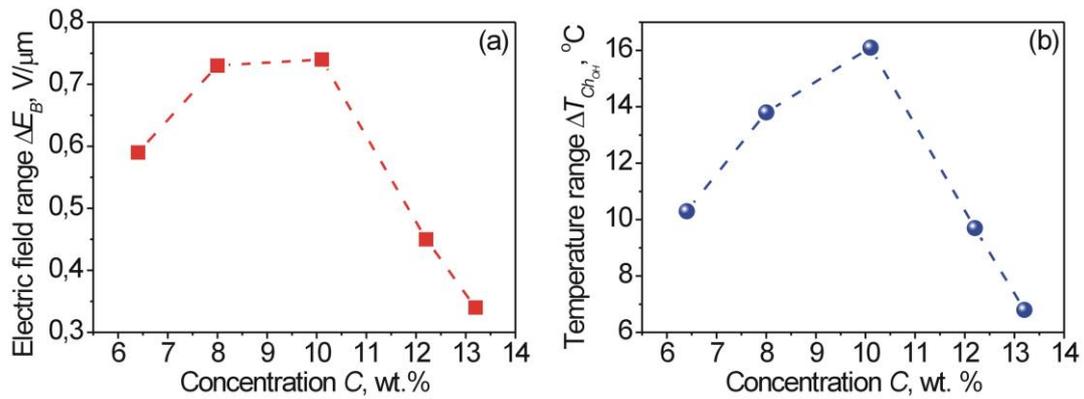

Figure 10. Dependence of the critical electric field range $\Delta E_B$ at 34 °C (a) and temperature range of the $Ch_{OH}$ state $\Delta T_{ChOH}$ (b) on concentration of ChD-3816 in the studied oblique helicoidal cholesteric

The photographs of the 20.5 μm LC cell filled by the $N^*_{tb}$-forming mixture with 10.1 wt. % of ChD-3816 in the $Ch_{OH}$ state (with selective Bragg reflection in the visible spectrum) are shown in Figure 11 (a)-(c) for various values of electric field $E$ ($f = 1$ kHz) .

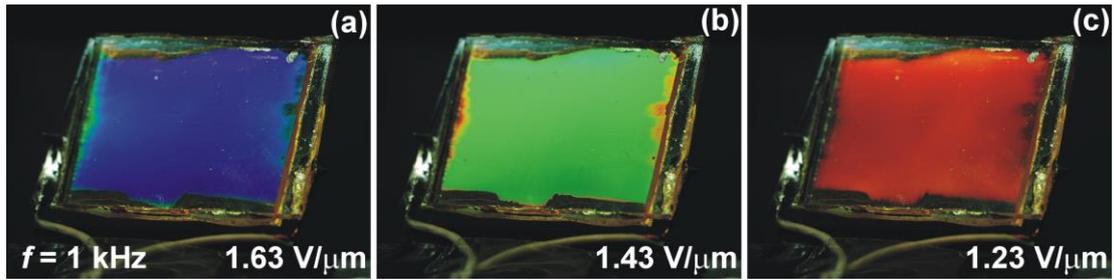

Figure 11. Photographs of the 22.1 μm LC cell with Ch$_{OH}$ formed under electric field: (a) 1.63 V/μm, (b) 1.43 V/μm and (c) 1.23 V/μm.

Figure 12 (a) shows the dynamics of the transmission spectrum of the Ch$_{OH}$ state recorded by spectrometer Ocean Optics USB4000. The sequential shifting of the selective reflection wavelength towards the range with longer wavelengths (*i.e.* so-called red shift) under decreasing electric field is shown in Figure 12 (b).

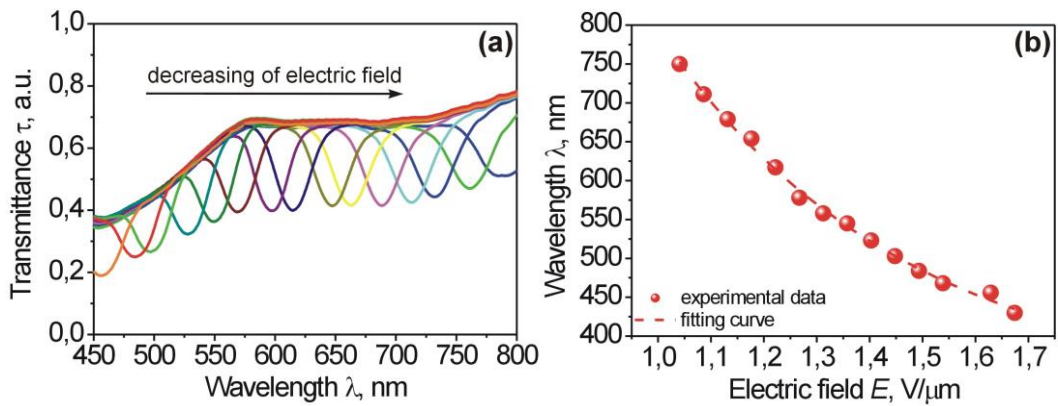

Figure12. (a) Dynamics of the transmission spectrum of the Ch$_{OH}$ under decreasing of the applied electric field *E*. (b) The dependence of the minimum of transmission spectrum of the Ch$_{OH}$ in the visible light range on the applied electric field *E* of frequency *f* = 1 kHz. LC cell was filled by N$^*_{tb}$ mixture consisting of 10.1 wt. % ChD-3816. The temperature of the Ch$_{OH}$ was 27.5 °C. The viewed angle of the LC cell was 45°.

### 3.2. Phase transitions of UV-irradiated N$^*_{tb}$

In this section, we will consider the effects of UV irradiation on the phase transition temperatures of the N$^*_{tb}$-forming mixtures. As chiral molecules of ChD-3816 are photosensitive, the transformation of the rod-like *trans*-isomer to a bended *cis*-isomer with a dramatic change in molecular geometry was expected to cause the change in the HTPs of these isomers. This, in turn, enables us, on the one hand, to tune the length of cholesteric pitch $P_0$ by the light exposure and, on the other hand, to change the order parameters by varying the temperatures of the phase transitions.

Figure 13 shows the dependence of the cholesteric pitch $P_0$ of the N* phase of the N*$_{tb}$-forming mixture on the irradiation time $t_{irr}$ (with no applied voltage). The gradual unwinding of cholesteric helix with irradiation time $t_{irr}$ was observed, with a tendency to saturation (as it should be expected from the known data on similar systems [34,35]).

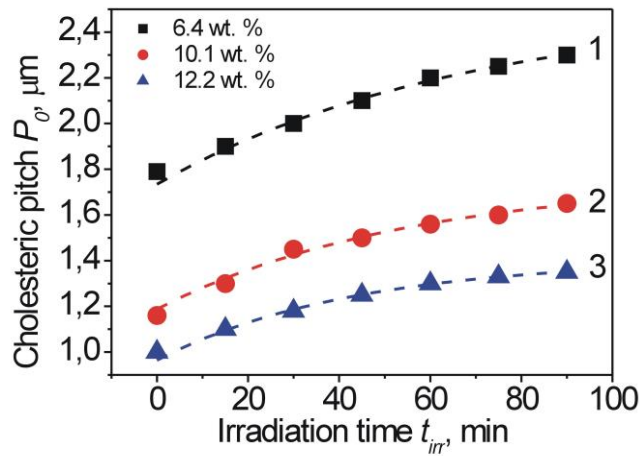

Figure 13. Helical pitch $P_0$ as function of the irradiation time $t_{irr}$ in the N* phase of the N*$_{tb}$-forming mixture at different concentrations of ChD-3816: (1) - 6.4 wt. % (solid black squares), (2) - 10.1 wt. % (solid red circles) and (3) - 12.2 wt. % (solid blue triangles).

Under UV irradiation, the unwinding of the cholesteric helix leads to increasing of the value of the electric field $E$ required to switch on the selective Bragg reflection in the visible range. Figure 14 (a) shows the dependence of the critical electric field $E_B$ on irradiation time $t_{irr}$ for various concentrations $C$ of the ChD-3816, namely 6.4 wt. % (red circles), 10.1 wt. % (green squares) and 12.2 wt. % (blue triangles) for $\lambda_1 = 430$ nm (solid symbols) and $\lambda_2 = 720$ nm (opened symbols).

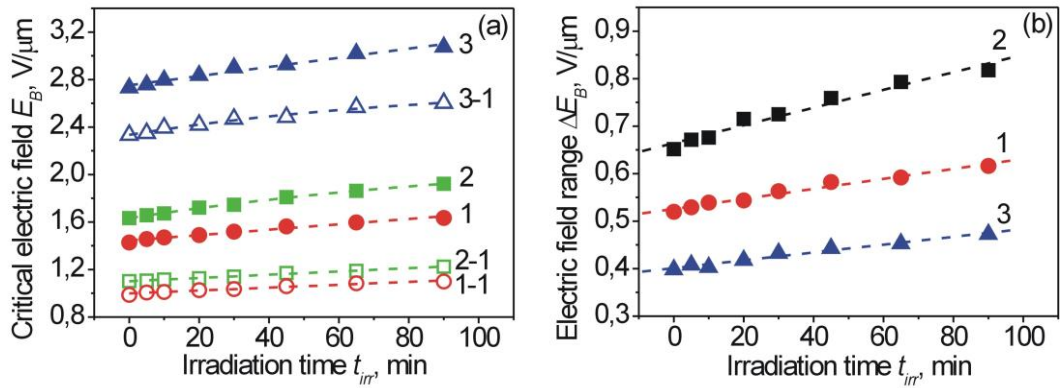

Figure 14. Dependence of the critical electric field $E_B$ (a) and width of the electric field range $\Delta E_B$ (b) on irradiation $t_{irr}$ for the basic $N_{tb}$-forming mixture doped by ChD-3816 with various concentrations: (1) – 6.4 wt. % (red circles), (2) – 10.1 wt. % (green squares) and (3) – 12.2 wt. % (blue triangles). (a) The critical electric field $E_B$ was measured at $\lambda_1 = 430$ nm (closed symbols) and $\lambda_2 = 720$ nm (opened symbols). (b) The width of the electric field range $\Delta E_B$ was calculated as difference between the critical electric field $E_B$ measured at 430 nm and 720 nm for ChD-3816 concentrations: 6.4 wt. % (red circles, line 1), 10.1 wt. % (black squares, line 2) and 12.2 wt. % (blue triangles, line 3).

Figure 14 (b) shows the electric field range $\Delta E_B$ (i.e. $\Delta E_B = E_B^{\lambda_1=430} - E_B^{\lambda_2=720}$) widened with prolonged UV exposure of the studied system. The prolonged UV illumination leads to broader $\Delta E_B$, which depends non-monotonously on the ChD-3816 concentration. The broadest $\Delta E_B$ range is observed for $C = 10.1$ wt. %, as can be seen from Figure 14 (b).

Figure 15 shows the effect of the irradiation time $t_{irr}$ on temperatures of phase transitions of the studied $N^*_{tb}$-forming systems with different concentrations of the light-sensitive ChD-3816. The increasing of the irradiation time $t_{irr}$ leads to monotonous decreasing of temperatures of phase transitions for both the $N^*_{tb} - N^*$ (Figure 15 (a)) and the $N^* -$ Iso (Figure 15 (b)), which is caused by *trans-cis* isomerisation of ChD-3816 molecules.

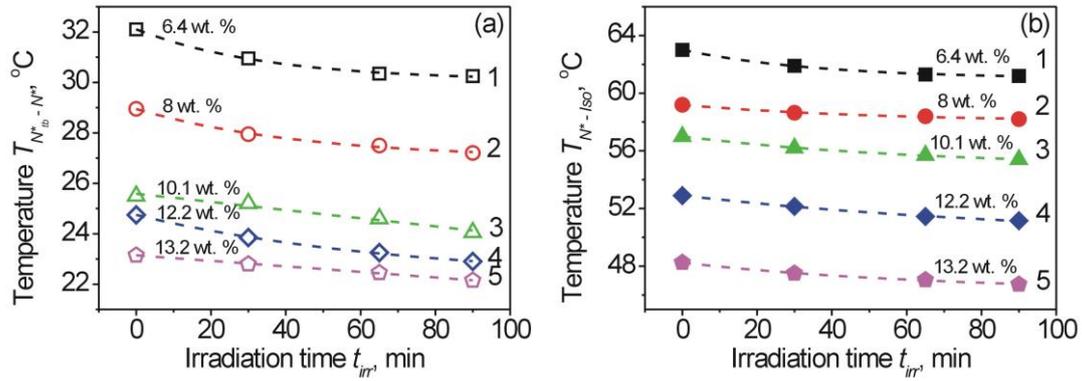

Figure 15. Dependence of the average temperatures of the phase transitions of $N^*_{tb}$ (*e.g.* during heating and cooling process) on irradiation time $t_{irr}$ for: (a) $N^*_{tb}$ to $N^*$ (opened symbols) and (b) $N^*$ to Iso (solid symbols). The basic $N^*_{tb}$ contains various concentrations of the light-sensitive ChD-3816, namely: (1) - 6.4 wt. % (black squares), (2) - 8 wt. % (red circles), (3) - 10.1 wt. % (green triangles), (4) - 12.2 wt. % (blue diamonds) and (5) - 13.2 wt. % (magenta pentagons). Irradiation of $N^*_{tb}$ was carried out by UV lamp with $\lambda_{max}$ = 365 nm for various irradiation time $t_{irr}$: 30, 65 and 90 min. The rates of the temperature change during the heating/cooling process was 0.1 °C/min.

In the non-irradiated area the concentration of *trans*-isomer of the ChD-3816 is higher in comparison with the irradiated area. Due to this, the temperatures of phase transition $N^*_{tb} - N^*$ are different in these areas. The appearance of selective Bragg reflection of the $Ch_{OH}$ state in electric field $E$ at the temperature near to phase transition $N^*_{tb} - N^*$ will be observed at lower temperature in comparison with non-irradiated area. This can be used to obtain high contrast at the boundary between the irradiated and non-irradiated areas (Figure 16).

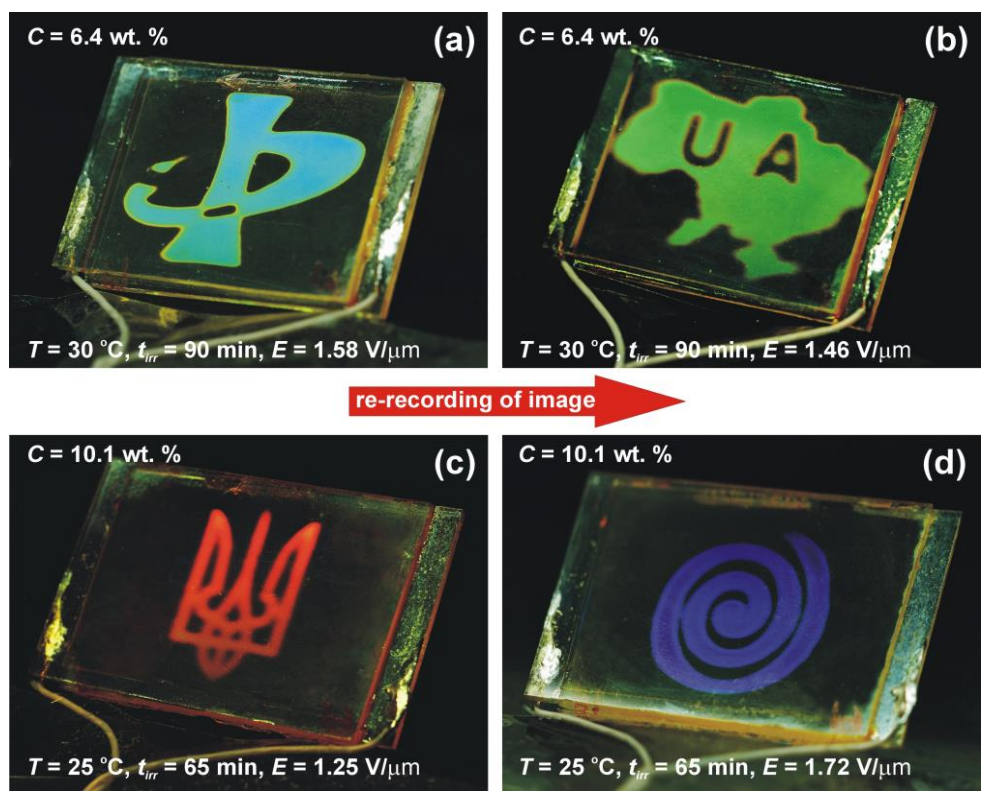

Figure 16. Photographs of samples with high contrast images recorded by UV irradiation for N*$_{tb}$-forming systems with different concentration of ChD-3816.

Photos of samples with the highest contrast of the images recorded by UV irradiation are shown in Figure 16. Owing to the reversible *trans-cis* isomerisation of ChD-3816 molecules, the N*$_{tb}$ phase, possessing Ch$_{OH}$ state in wide temperature and electrically controllable ranges, can be used to re-recording of information (*e.g.* Figure 16 (a), (b) and Figure 16 (c), (d)).

**Conclusion**

In this manuscript we studied light-sensitive chiral nematic twist-bend phase N*$_{tb}$, consisting of basic nematic twist-bend N$_{tb}$ mixture (*i.e.* two twist-bend nematic phases of achiral liquid crystals dimers, namely 1',7"-bis-4-(4-cyanobiphen-4'-yl)heptane (CB7CB), 1-(4-cyanobiphenyl-4'-yloxy)-6-(4-cyanobiphenyl-4'-yl)hexane (CB6OCB) and monomer nematic 4-pentyl-4'-cyanobiphenyl (K15 or 5CB) in the ratio of 39:19:42) doped by a newly synthesized light-sensitive chiral dopant (1*R*,2*S*,5*R*)-2-isopropyl-5-methylcyclohexyl-4-{(*E*)-[4-(hexanoyloxy)phenyl]diazenyl}benzoate (ChD-3816). The concentration dependence and influence of UV irradiation on temperatures of phase transitions were studied. The decrease of the phase transition temperatures both with increasing concentration the ChD-3816 and irradiation time was found. At certain concentrations of the ChD-3816 there exists a wide temperature interval of N* phase where oblique heliconical cholesteric state with selective Bragg reflection in the visible light spectrum can be observed. The critical electric field required to switch on the selective reflection in the visible range increases with increasing of the temperature of N* phase and irradiation time. It has been also shown how the use of light-sensitive chiral dopants can significantly expand the possibilities of wavelength switching not only by means of the applied electric field but also using the light, thus opening prospects of new "photo-electro-optical" effects in liquid crystals.

Acknowledgements

The authors thank W. Becker (Merck, Darmstadt, Germany) for his generous gift of nematic liquid crystals E7. Dr. A. Buchnev (Cambridge, England) who had supplied us with polymer PI2555, and Field service specialist IV V. Danylyuk (Dish LLC, USA) for his gift of some Laboratory equipment. Mesogens CB7CB and CB6OCB were generously donated by the Materials and Manufacturing Directorate of the Air Force Research Laboratory, Wright-Patterson AFB, Ohio. The authors thank Prof. A. Tolmachev (Chemico-Biological Center Taras Shevchenko National University of Kyiv) and Dr. S. Lukyanets (Institute of Physics, NAS of Ukraine) for the helpful discussions. This study was supported by the National Academy of Medical Sciences of Ukraine within project No. 0120U101532 and the National Academy of Sciences of Ukraine within project No. 0122U002636.

Supporting Information

# Light-sensitive oblique heliconical cholesteric based on $N_{tb}$ phase doped by chiral azo-compound

Vitalii Chornous[a], Alina Grozav[a], Myhaylo Vovk[b], Daria Bratova[c,d], Natalia Kasian[e], Longin Lisetski[e], Igor Gvozdovskyy[d]*

## S1. Synthesis and chemical characterization of ChD-3816

For the synthesis of the **(1$R$,2$S$,5$R$)-2-isopropyl-5-methylcyclohexyl-4-{($E$)-[4-(hexanoyloxy)phenyl]diazenyl}benzoate** (or, for short, ChD-3816) the initial materials and reagents were used from Enamine Ltd (Kyiv, Ukraine); the "p.a." grade solvents were used.

The NMR spectra were recorded with Varian VXR-400(500) instrument (400 MHz for $^1$H and 125.7, 150.8 MHz for $^{13}$C) in DMSO-$d_6$ and CDCl$_3$ solutions, with TMS as an internal standard. Chemical shifts (δ) and J values of spectra are given in ppm and Hz, respectively. Structures of spectral lines are designed as: s (singlet), d (doublet), t (triplet), td (triplet of doublet) and m (multiplet).

LC-MS spectra were recorded by means of an Agilent 1100 Series high performance liquid chromatograph (Hewlett-Packard, California, USA) equipped with a diode matrix with an Agilent LC\MSD SL mass selective detector. The approximate values of melting points were determined by a Kofler bench.

The synthesis of (1$R$,2$S$,5$R$)-2-isopropyl-5-methylcyclohexyl 4-{(E)-[4-(hexanoyloxy)phenyl] diazenyl} benzoate (ChD-3816 for short) was performed according to the following scheme:

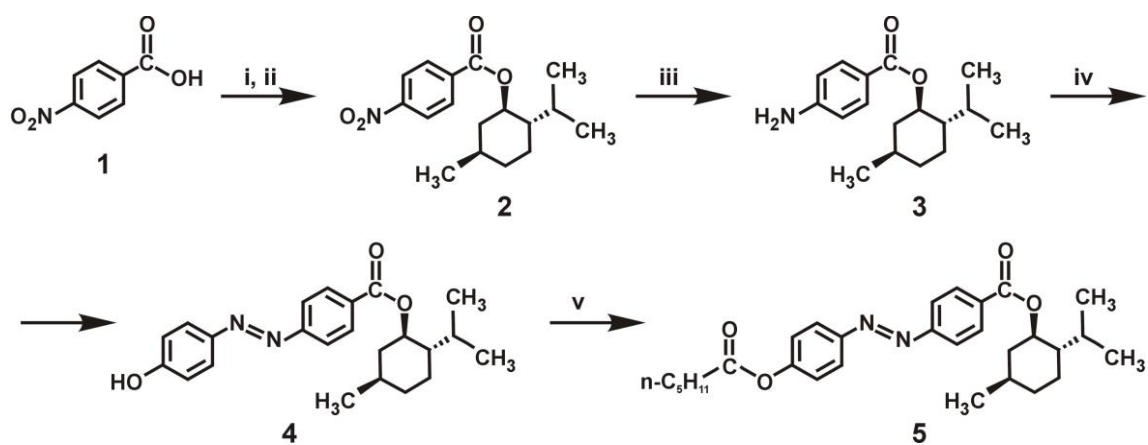

Figure 1S1. Synthesis of (1*R*,2*S*,5*R*)-2-isopropyl-5-methylcyclohexyl 4-{(E)-[4-(hexanoyloxy)phenyl] diazenyl} benzoate (**5**) by using the reagents and conditions for various stages: (i) - toluene, SOCl$_2$, reflux; (ii) - *l*-menthol, Et$_3$N, acetonitrile, reflux; (iii) - SnCl$_2$·2H$_2$O, ethanol, reflux; (iv) - NaNO$_2$, HCl, C$_6$H$_5$OH, 0-5 °C; and (v) - n-C$_5$H$_{11}$COCl, Et$_3$N, acetonitrile, reflux.

**(1*R*,2*S*,5*R*)-2-Isopropyl-5-methylcyclohexyl-4-nitrobenzoate (compound 2).** 16.7 g (0.1 mol) of 4-nitrobenzoic acid, 100 ml of dry toluene and 1 drop of DMF were added to a 500 ml reactor equipped with a magnetic stirrer. 17.9 g (0.15 mol) of SOCl$_2$ were added by dropwise to the reaction mixture with stirring and external cooling. After the reaction mixture was heating under reflux within 2 hours to complete gas evolution, and further the solvent was evaporated under reduced pressure. The residue was dissolved in 100 ml of acetonitrile and 15.6 (0.1 mol) of *l*-menthol was added, and to the resulting mixture 13.3 g (0.13 mol) of triethylamine was added by dropwise with stirring and external cooling. The mixture was heating under reflux within 2 hours, and the solvent was evaporated under reduced pressure and washed with water. The compound **2** was crystallized from ethanol.

M. p. 64 - 65 °C. **$^1$H-NMR** (400MHz, DMSO-*d*$_6$): δ 8.33 (d, *J* = 8.8 Hz, 2H, ArH), 8.16 (d, *J* = 8.8 Hz, 2H, ArH), 4.86 (td, *J$^1$* = 4.0 Hz, *J$^2$* 6.8 Hz, 1H, OCH), 2.00-1.98 (m, 1H, CH), 1.87-1.81 (m, 1H, CH), 1.66-1.64 (m, 2H, CH$_2$), 1.55-1.49 (m, 2H, CH$_2$), 1.13-1.03 (m, 2H, CH$_2$), 0.89-0.85 (m, 7H, 2CH$_3$+CH), 0.72 (d, *J* = 6.8 Hz, 3H, CH$_3$).

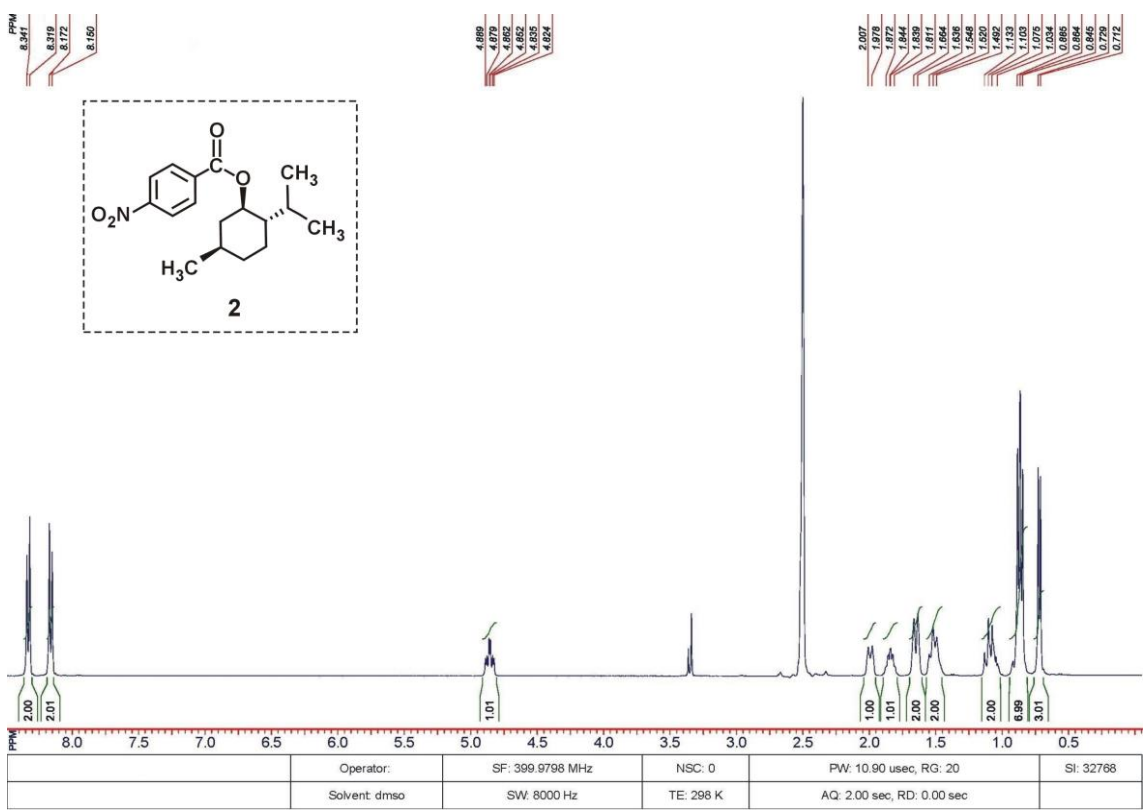

Figure 2S1. ¹H-NMR spectrum of the (1*R*,2*S*,5*R*)-2-Isopropyl-5-methylcyclohexyl-4-nitrobenzoate (compound **2**).

**¹³C-NMR** (151 MHz, DMSO-*d*₆): δ 164.2 (C = O), 150.7, 135.7, 130.9, 124.3, 75.7, 46.6, 40.8, 34.1, 31.3, 26.6, 23.6, 22.3, 20.9, 16.8. LC-MS, *m/z* (%): 306(100) [M+1]. Anal. Calcd for $C_{17}H_{23}NO_4$, (305.38): C, 66.86 %; H, 7.59 %; N, 4.59 %; O, 20.96 %. Found: C, 66.76 %; H, 7.47 %; N, 4.65 %; O, 21.08 %.

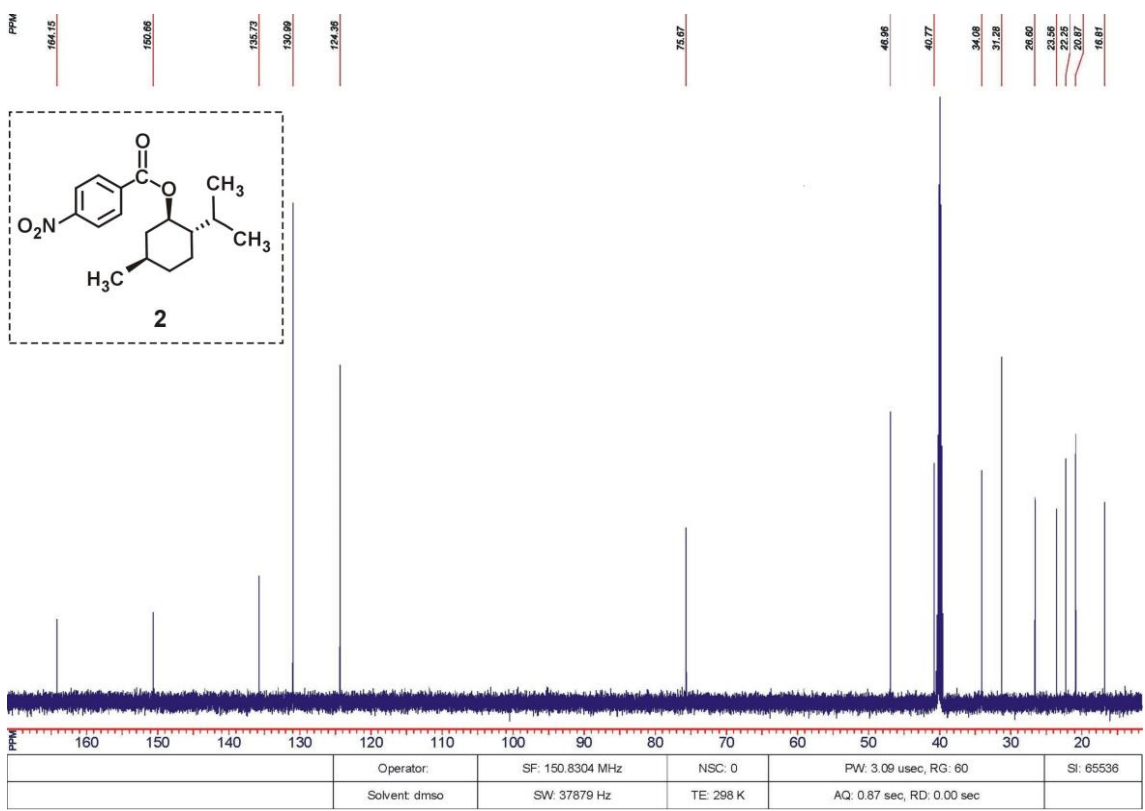

Figure 3S1. $^{13}$C-NMR spectrum of the (1$R$,2$S$,5$R$)-2-Isopropyl-5-methylcyclohexyl-4-nitrobenzoate (compound **2**).

**(1$R$,2$S$,5$R$)-2-Isopropyl-5-methylcyclohexyl-4-aminobenzoate (compound 3).** In a 500 ml single-necked flask was added 100 ml of ethanol, 15 g (0.05 mol) of compound **2** and 56.5 g (0.25 mol) of SnCl$_2$·2H$_2$O under argon. The reaction mixture was heating under reflux for one hour and the solvent was evaporated under reduced pressure. The residue was dissolved in ethyl acetate and washed with 30 % NaOH solution (3 × 20 ml). The organic layer was separated and dried over anhydrous Na$_2$SO$_4$. After evaporation of the solvent in a vacuum, the residue was purified by silica gel column chromatography by using hexane as eluent.

M.p. 90 – 91 ºC. **$^1$H-NMR** (400 MHz, DMSO-$d_6$): δ 7.62 (d, $J$ = 8.4 Hz, 2H, ArH), 6.56 (d, $J$ = 8.4 Hz, 2H, ArH), 5.93 (s, 1H, NH), 4.83 (td, $J^1$ = 4.4 Hz, $J^2$ = 6.4 Hz, 1H, OCH), 1.95-1.92 (m, 1H, CH), 1.87-1.83 (m, 1H, CH), 1.66-1.63 (m, 2H, CH$_2$), 1.48-1.43 (m, 2H, CH$_2$), 1.09-0.99 (m, 2H, CH$_2$), 0.92-0.85 (m, 7H, 2CH$_3$+CH), 0.72 (d, $J$ = 7.2 Hz, 3H, CH$_3$).

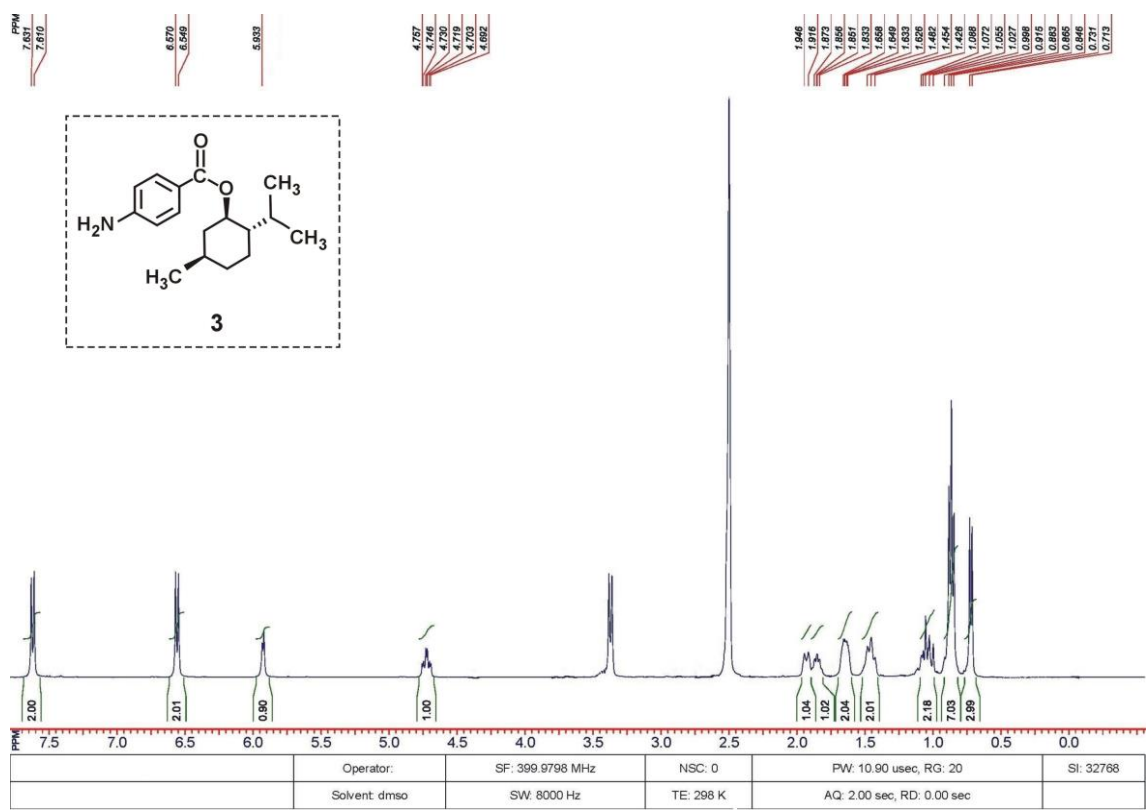

Figure 4S1. $^1$H-NMR spectrum of the (1$R$,2$S$,5$R$)-2-Isopropyl-5-methylcyclohexyl-4-aminobenzoate (compound **3**).

$^{13}$**C-NMR** (125.7 MHz, DMSO- $d_6$): δ 165.4 (C = O), 153.4, 131.0, 116.3, 112.6, 72.7, 46.8, 40.9, 33.9, 30.9, 26.2, 23.4, 21.9, 20.5, 16.6. LC-MS, *m/z* (%): 276(100) [M+1]. Anal. Calcd for C$_{17}$H$_{25}$NO$_2$ (275.39): C, 74.14%; H, 9.15%; N, 5.09%; O, 11.62%. Found: C, 74.23%; H, 9.03 %; N, 5.18 %; O, 11.53 %.

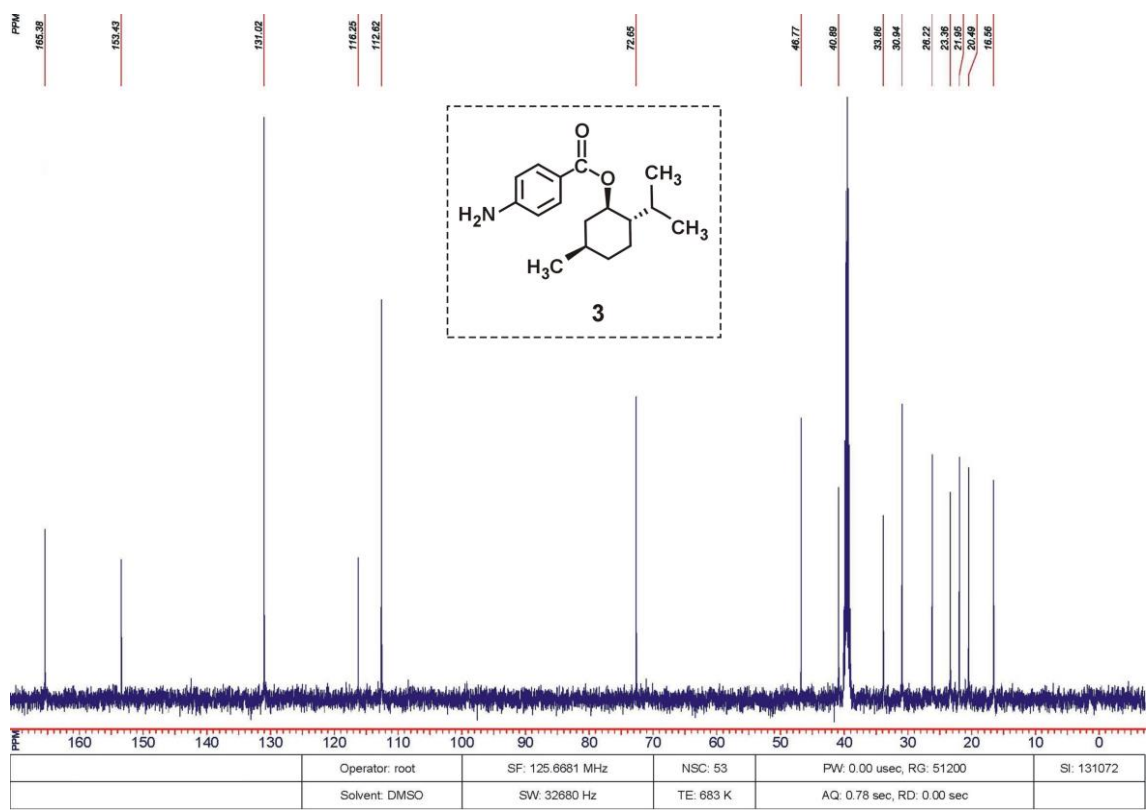

Figure 5S1. $^{13}$C-NMR spectrum of the (1R,2S,5R)-2-Isopropyl-5-methylcyclohexyl-4-aminobenzoate (compound **3**).

**(1R,2S,5R)-2-isopropyl-5-methylcyclohexyl-4-[(E)-(4-hydroxyphenyl)diazenyl]benzoate (compound 4).** To a solution of 2.75 g (0.01 mol) of compound **3** in 20 ml of CH$_3$COOH was added 6 g (0.05 mol) of 30 % HCl and 10 g of crushed ice. A solution of 0.78 g (0.011 mol) of sodium nitrite in 3 ml of water was added dropwise to the resulting mixture with stirring. After 15 min, the resulting diazonium salt solution was added to a solution obtained by dissolving 1 g (0.011 mol) of phenol and 2.8 g (0.07 mol) of sodium hydroxide in 50 ml of water. The red precipitate was filtered off, dried and purified by chromatography on silica gel by using mixture of the eluent hexane:ethyl acetate, 1:1. The compound **4** was as yellow-red crystals.

M.p. 97-98 °C. **$^1$H-NMR** (400 MHz, DMSO-$d_6$): δ 10.5 (s, 1H, OH), 8.10 (d, $J$ = 8.4 Hz, 2H, ArH), 7.90-7.85 (m, 4H, ArH), 6.97 (d, $J$ = 8.4 Hz, 2H, ArH), 4.87-4.82 (m, 1H, OCH), 2.00-1.97 (m, 1H, CH), 1.88-1.85 (m, 1H, CH), 1.66-1.63 (m, 2H, CH$_2$), 1.54-1.49 (m, 2H, CH$_2$), 1.13-1.04 (m, 2H, CH$_2$), 0.88-0.85 (m, 7H, 2CH$_3$ +CH), 0.73 (d, $J$ = 6.8 Hz, 3H, CH$_3$).

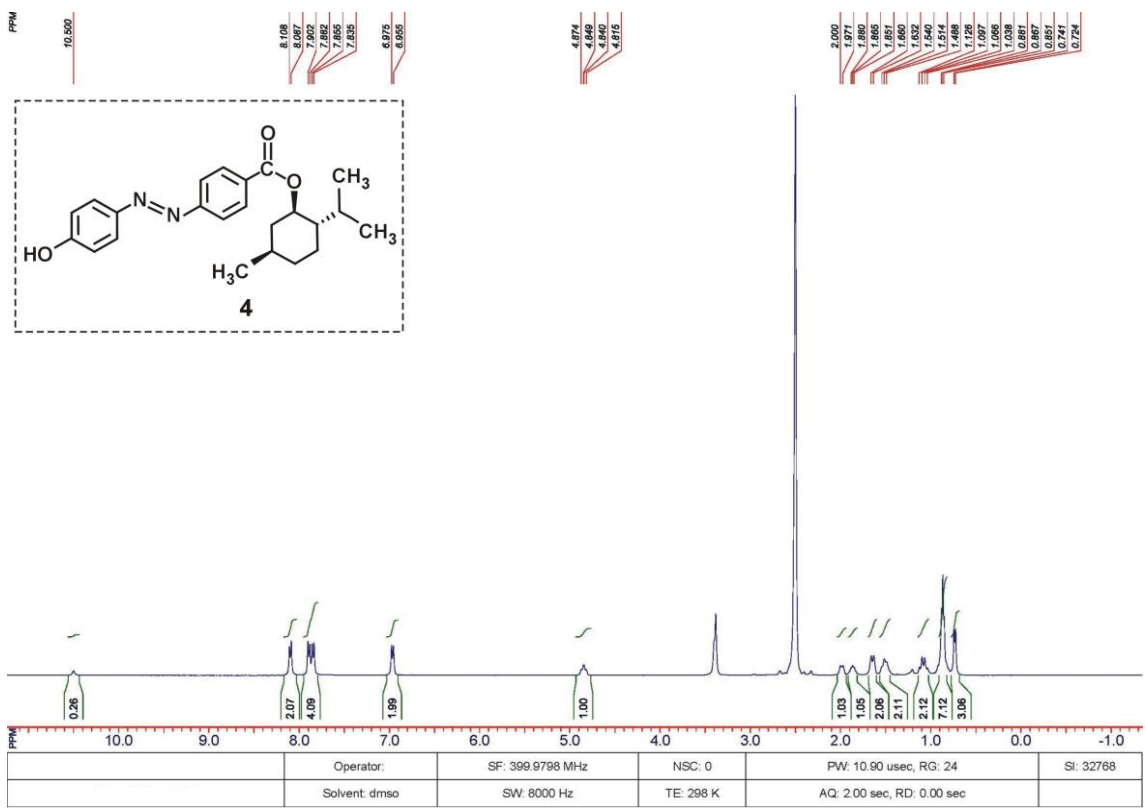

Figure 6S1. ¹H-NMR spectrum of the (1*R*,2*S*,5*R*)-2-isopropyl-5-methylcyclohexyl-4-[(*E*)-(4-hydroxyphenyl)diazenyl]benzoate (compound **4**).

**¹³C-NMR** (150.8 MHz, DMSO-*d*₆): δ 165.1(C = O), 155.2, 145.8, 131.4, 130.8, 125.8 (2C), 122.7, 116.4, 74.8, 46.9, 40.9, 34.1, 31.3, 26.6, 23.6, 22.2, 20.8, 16.9. LC-MS, *m/z* (%): 381(100) [M+1]. Anal. Calcd for $C_{23}H_{28}N_2O_3$ (380.49): C, 72.61%; H, 7.42%; N, 7.36%; O, 12.61%. Found: C, 72.48%; H, 7.34%; N, 7.47%; O, 12.54%.

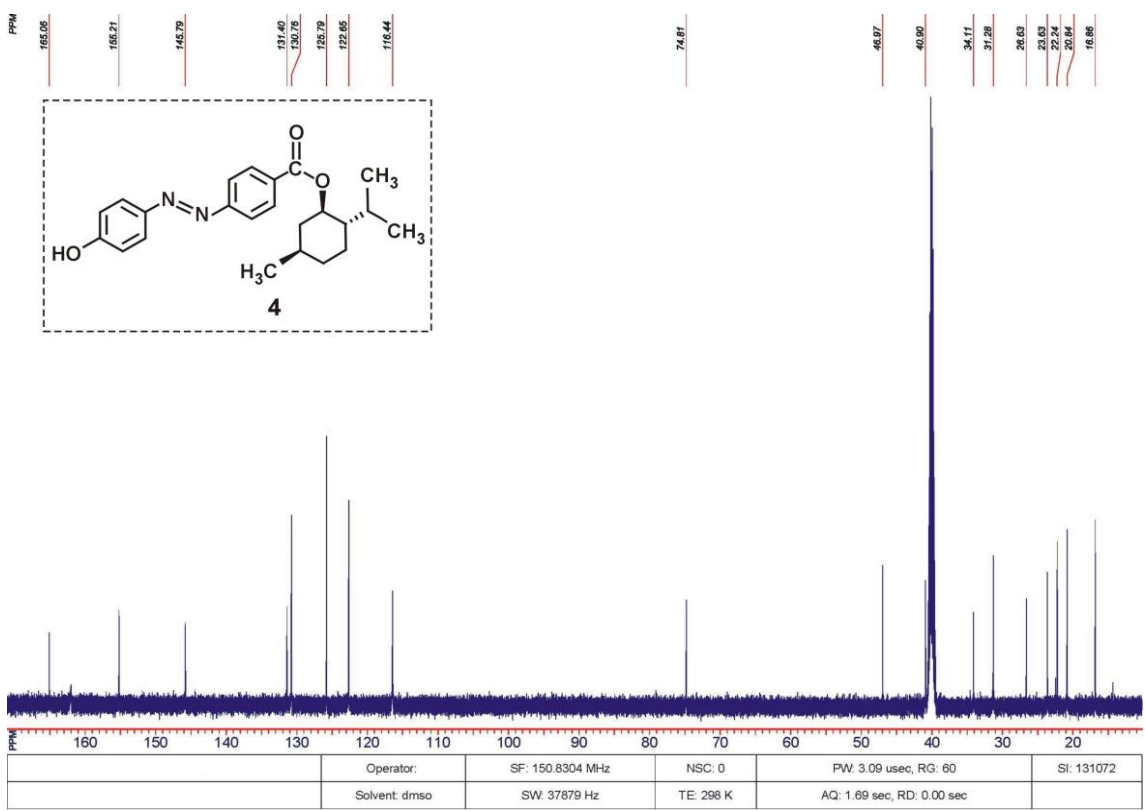

Figure 7S1. $^{13}$C-NMR spectrum of the (1$R$,2$S$,5$R$)-2-isopropyl-5-methylcyclohexyl-4-[($E$)-(4-hydroxyphenyl)diazenyl] benzoate (compound **4**).

**(1$R$,2$S$,5$R$)-2-isopropyl-5-methylcyclohexyl-4-{($E$)-[4-(hexanoyloxy)phenyl]diazenyl}benzoate 5 (ChD-3816).** To solution of 11.4 g (0.03 mol) of compound **4** and 3 g (0.03 mol) of triethylamine in 50 ml of acetonitrile was added 4 g (0.03 mol) of hexanoyl chloride. The reaction mixture was heating under reflux within 2 hours; the solvent was evaporated under reduced pressure. The red precipitate was filtered off, dried and purified by chromatography on silica gel. It was used hexane as eluent. The compound **5** (ChD-3816) was as yellow plates.

M.p.65.7 °C. **$^1$H-NMR** (400 MHz, CDCl$_3$): δ 8.21 (d, $J$ = 8.4 Hz, 2H, ArH), 8.02-7.94 (m, 4H, ArH), 7.28 (d, $J$ = 8.0 Hz, 2H, ArH), 4.99 (td, $J^1$ = 4.2 Hz, $J^2$ = 6.4 Hz, 1H, OCH ), 2.61(t, $J$ = 8.0 Hz, 2H, CH$_2$COO), 2.19-2.16 (m, 1H, CH), 2.02-1.98 (m, 1H, CH), 1.82-1.75 (m, 4H, CH$_3$+CH), 1.63-1.58 (m, 2H, CH$_2$), 1.45-1.42 (m, 4H, 2CH$_2$), 1.21-1.11 (m, 2H, CH$_2$), 0.97-0.95 (m, 10H, 2CH$_3$+2CH$_2$), 0.83 (d, $J$ = 6.8 Hz, 3H, CH$_3$).

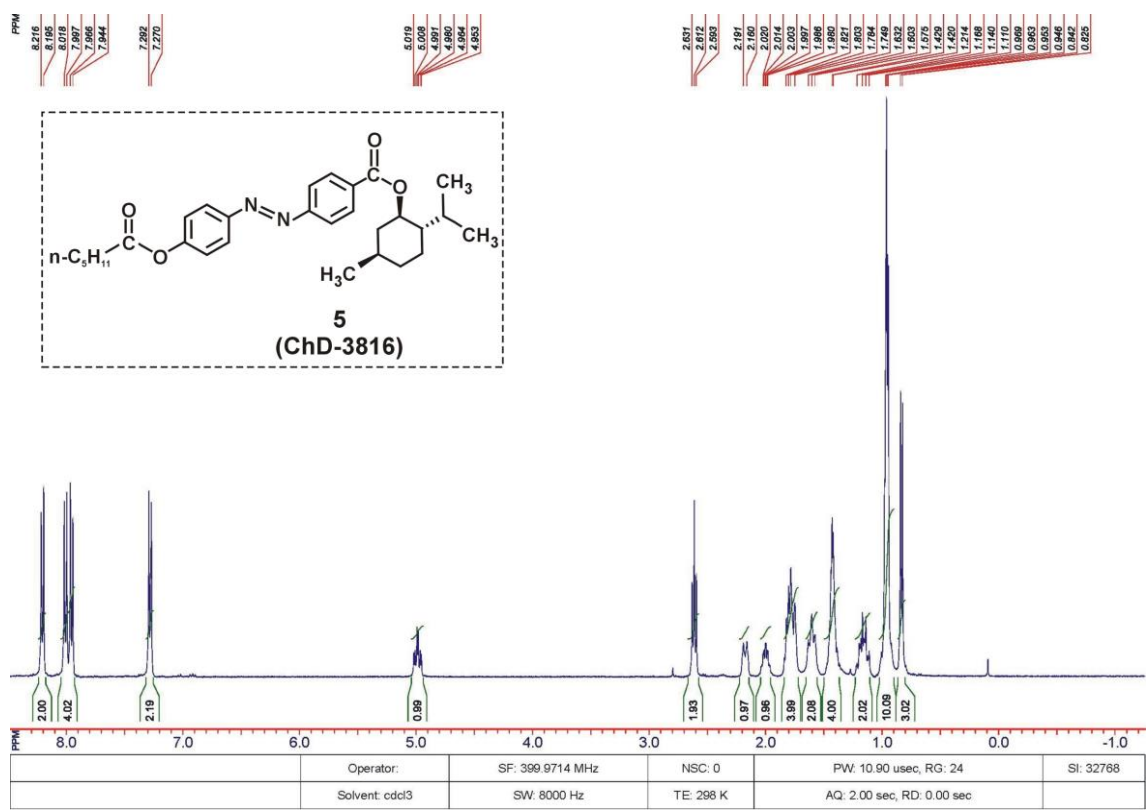

Figure 8S1. ¹H-NMR spectrum of the (1*R*,2*S*,5*R*)-2-isopropyl-5-methylcyclohexyl-4-{(*E*)-[4-(hexanoyloxy)phenyl]diazenyl}benzoate (compound **5** or, for short, ChD-3816).

**¹³C-NMR** (150.8 MHz, DMSO-$d_6$): δ 171.90 (C = O), 165.0 (C = O), 154.8, 153.8, 149.9, 132.4, 130.9, 124.6, 123.4, 123.2, 75.0, 47.0, 40.9, 34.1, 33.9, 31.3, 31.0, 26.7, 24.4, 23.7, 22.3, 22.2, 20.9, 16.9, 14.2. LC-MS, *m/z* (%): 479 (100) [M+1]. Anal. Calcd for $C_{29}H_{38}N_2O_4$ (478.64): C, 72.77%; H, 8.00%; N, 5.85%; O, 13.37%. Found: C, 72.85 %; H, 8.11 %; N, 5.93 %; O, 13.21 %.

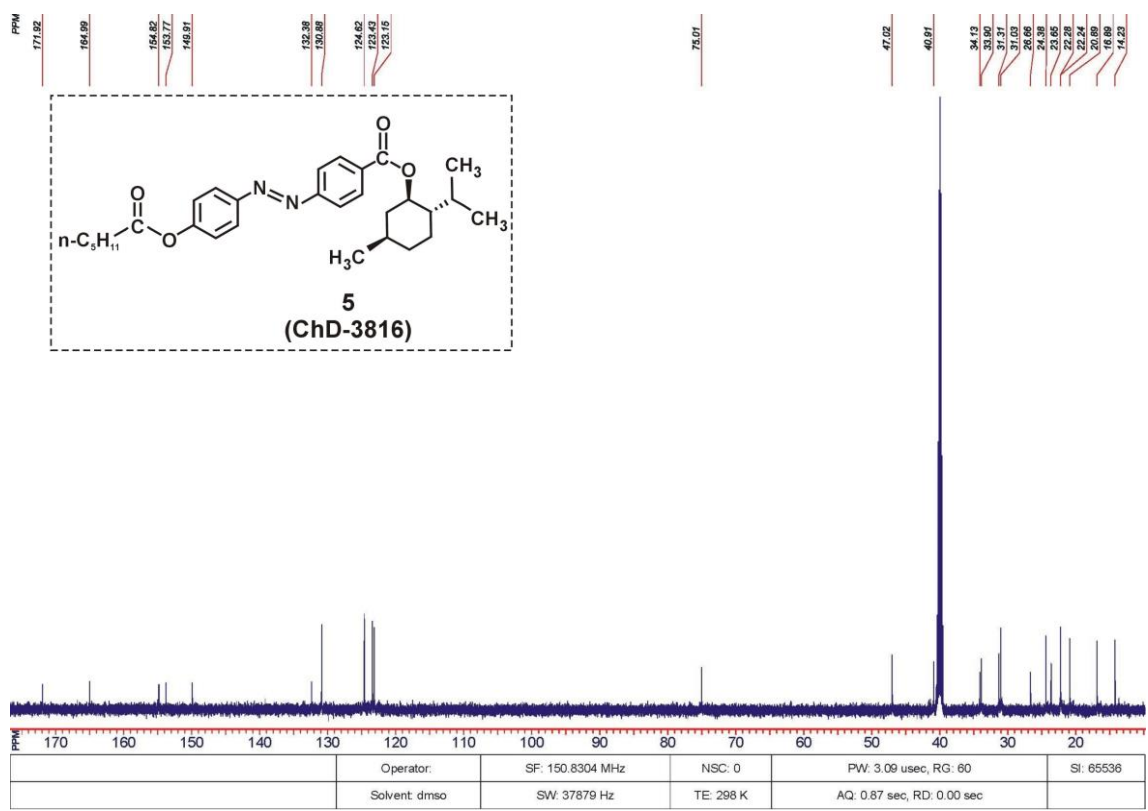

Figure 9S1. $^{13}$C-NMR spectrum of the (1*R*,2*S*,5*R*)-2-isopropyl-5-methylcyclohexyl-4-{(*E*)-[4-(hexanoyloxy)phenyl]diazenyl}benzoate (compound **5** or , for short, ChD-3816).

## S2. *Trans-cis* isomerisation of the ChD-3816 in ethanol solution

The compound ChD-3816 is a photosensitive chiral molecule possessing reversible *trans-cis* isomerisation under UV, vis and temperature. Figure 2S1 shows the scheme of the of the *trans-cis* and *cis-trans* isomerisations of the ChD-3501 molecule under UV, vis irradiation and temperature *ΔT*.

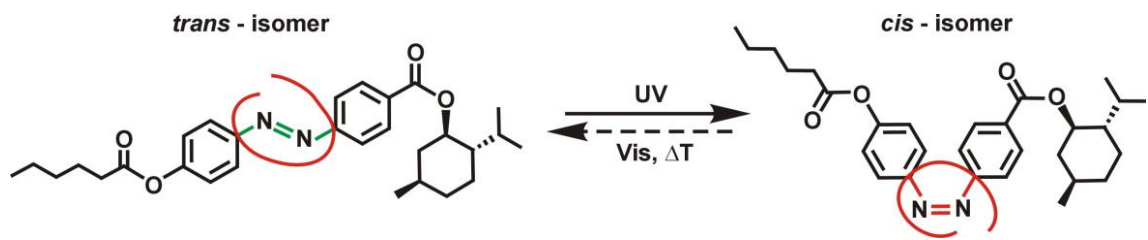

Figure 2S1. Scheme of the *trans-cis* and *cis-trans* isomerisations of the ChD-3501 molecule under UV, vis irradiation and temperature ΔT.

To study the effects of UV-vis radiation and temperature on the reversible *trans-cis* isomerisation (Figure 2S1), ChD- 3816 was dissolved in absolute ethanol (Enamine

Ltd). The ethanol solution of the ChD-3816 ($C$ = 0.03 mg/ml) possesses initial optical density OD ~ 1.7 in quartz cuvette with thickness 1 cm. The ethanol solution was of light orange colour.

Transformations of absorption spectrum of the ethanol solution of the ChD-3816 upon irradiation by UV lamp with $\lambda_{max}$ = 365 nm (*trans-cis* isomerisation) and incandescent lamp (*cis-trans* isomerisation) are shown in Figure 2S2.

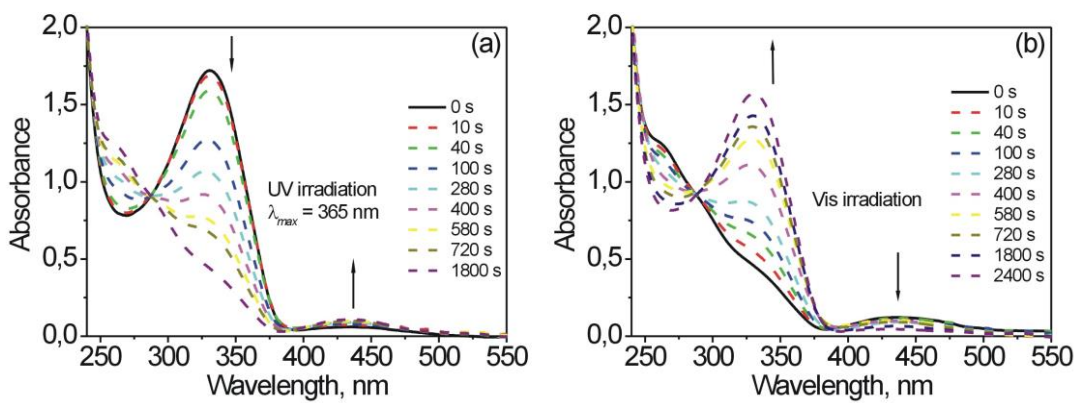

Figure 2S2. Transformations of absorption spectrum of the ethanol solution of the ChD-3816 upon irradiation by: (a) UV lamp with $\lambda_{max}$ = 365 nm (0 - 0 s, 1 - 10 s, 2 - 40 s, 3 - 100 s, 4 - 280 s, 5 - 400 s, 6 – 580s, 7 - 720 s, 8 – 1800s), (b) incandescent lamp (0 - 0 s, 1 - 10 s, 2 - 40 s, 3 – 100 s, 4 – 280 s, 5 – 400 s, 6 – 580 s, 7 – 720 s, 8 – 1800 s, 9 - 2400 s). Thickness of quartz cuvette is 1 cm.

Figure 2S3 shows transformations of absorption spectrum of the ethanol solution of the ChD-3816 previously irradiated by UV lamp with $\lambda_{max}$ = 365 nm within 1800 s (*trans-cis* isomerisation), during storage process of it at 80 °C (*cis-trans* isomerisation).

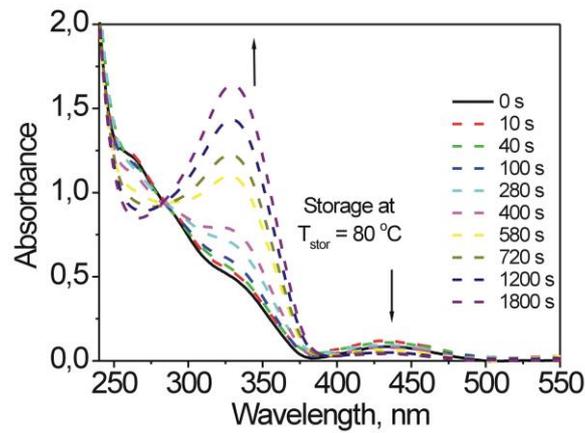

Figure 2S3. Transformations of absorption spectrum of the ethanol solution of ChD-3816 during storage at 80 °C (0 - 0 s, 1 - 10 s, 2 - 40 s, 3 - 100 s, 4 - 280 s, 5 - 400 s, 6 – 580s, 7 - 720 s, 8 – 1200 s, 9 – 1800s),

**S3. Cholesteric phase induced by ChD-3816**

Dissolution of ChD-3816 in nematic E7 (Merck, Darmstadt, Germany) leads to induction of the helicoidal structure possessing the helix with pitch that can be change in wide range of values (*i.e.* from nm to μm). Figure 3S1 shows the length of cholesteric pitch $P_0$ (a) and reciprocal cholesteric pitch $1/P_0$ (b) as function of concentration $C$ of ChD-3816.

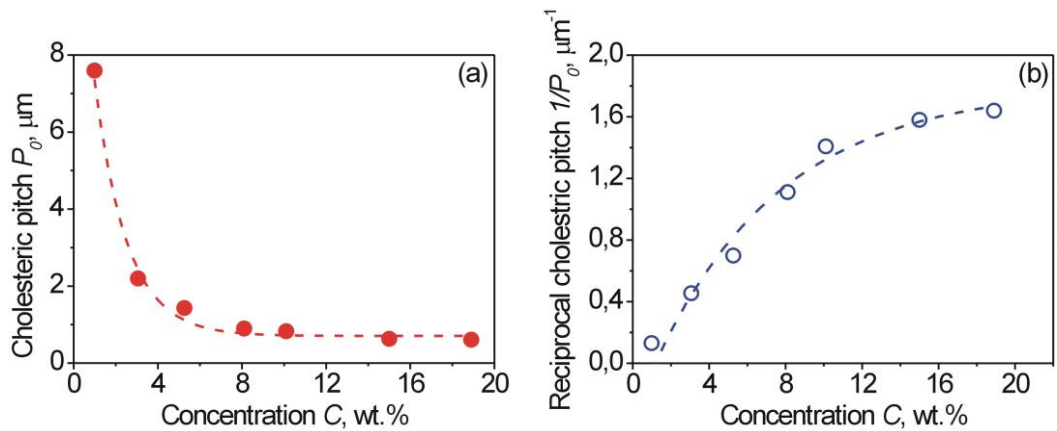

Figure 3S1. Dependence of the cholesteric pitch $P_0$ (a) and reciprocal cholesteric pitch $1/P_0$ (b) on concentration $C$ of the ChD-3816 in nematic E7.

To calculate the helical twisting power (HTP, $\beta$) of the ChD-3816 in nematic E7, we used the linear part of $1/P_0(C)$ plot, which should pass though the origin of coordinates (Figure 3S2).

It is known that the tangent of tilt angle of linear plot is the value of HTP of ChD dissolved in the nematic host [1]. The average value of HTP of the Ch-3816 is about 0.138 (μm × wt. %)$^{-1}$. The handedness of the cholesteric helix was measured by means of the effect of rotation of solid crystals of the ChD-3816 during the process of dissolution at the top of the nematic E7 droplet. The clockwise rotation of the solid crystal of chiral dopant ChD-3816 in POM was observed. As described elsewhere in detail [2-5], the clockwise rotation of small crystals of the various ChDs is typical for left-handed cholesteric helix induced by these chiral dopants. This is confirmed by the Grandjean-Cano method used to determine the helix handedness [6].

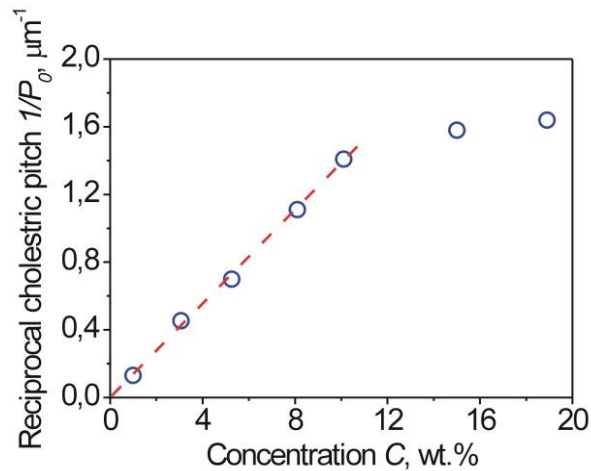

Figure 3S2. Linear part of the dependence $1/P_0(C)$ for ChD-3816 in E7.

**S4. UV irradiation of the induced cholesteric phase, based on ChD-3816 and nematic E7**

The irradiation of the cholesteric mixture, formed by 10 wt.% ChD-3816 added to the nematic E7 and filled into wedge-like LC cell, was carried out by UV lamp ($\lambda_{max}$ = 365 nm and power P = 6 W). The wedge-like LC cell was assembled from two glass plates covered by PI2555 (HD MicroSystems, USA) as aligning layer. PI2555 film was rubbed 15 times to obtain strong anchoring. The thickness of the thick end $d$ of the wedge-like LC cell was set as 5 μm by using Mylar spacers.

Figure 4S1 shows the change in the number of Grandjean-Cano lines ($N_{GC}$) in wedge-like LC cell during UV irradiation.

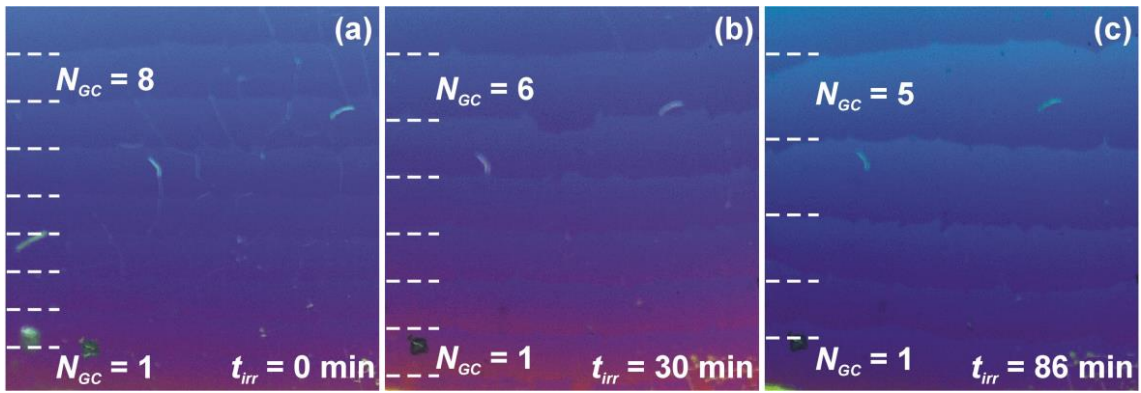

Figure 4S1. The change in $N_{GC}$ during irradiation by UV-lamp with $\lambda_{max}$ = 365 nm: (a) $N_{GC}$ = 8 at $t_{irr}$ = 0 min; (b) $N_{GC}$ = 6 at $t_{irr}$ = 30 min; and (c) $N_{GC}$ = 5 at $t_{irr}$ = 86 min.

Figure 4S2 shows the dependence of cholesteric helix pitch $P$ on irradiation time $t_{irr}$ during UV (a) and vis (b) illumination.

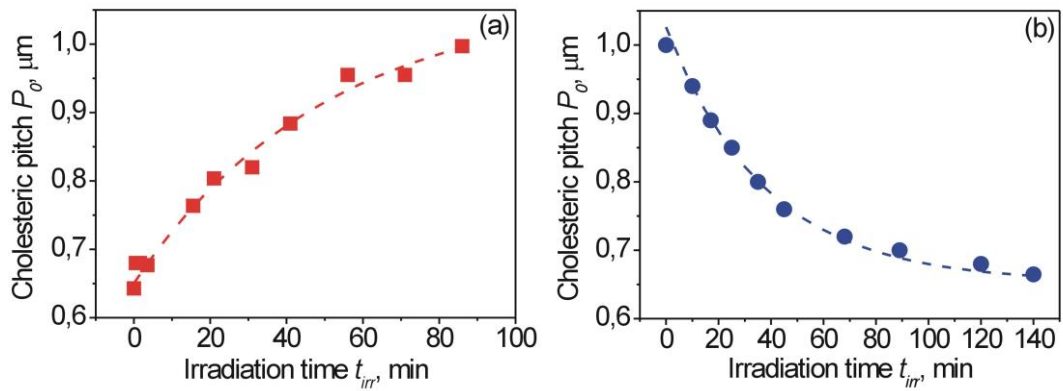

Figure 4S2. Dependence of cholesteric helix pitch $P_0$ on irradiation time $t_{irr}$ during illumination by: (a) UV-lamp with $\lambda_{max}$ = 365 nm and (b) incandescent lamp.